\documentclass[a4paper,10pt]{article}
\usepackage{graphicx}
\usepackage{amssymb}
\usepackage{amsmath}
\usepackage{amsfonts}
\usepackage{amsthm}
\usepackage{nicefrac}
\usepackage{dcolumn}
\usepackage{multirow}
\usepackage{cite}
\usepackage{mathrsfs}
\usepackage{bm}
\usepackage[usenames,dvipsnames]{xcolor}
\usepackage[colorlinks=true,citecolor=Blue,linkcolor=RubineRed,urlcolor=Blue]{hyperref}
\usepackage{bbm}
\usepackage{hyperref}
\usepackage{comment}
\usepackage{float}
\usepackage{caption}
\usepackage{subcaption}
\usepackage{xcolor}
\usepackage{soul}
\usepackage{tikz}

\setstcolor{red}

\begin{document}

\huge

\begin{center}
Electrical conductivities and low frequency opacities in the warm dense matter regime\end{center}

\vspace{0.5cm}

\large

\begin{center}
Mikael Tacu$^{a,b,}$\footnote{mikael.tacu@cea.fr}, Jean-Christophe Pain$^{a,b}$, Matthias Pautard$^a$ and Christophe Blancard$^{a,b}$
\end{center}

\normalsize

\begin{center}
\it $^a$CEA, DAM, DIF, F-91297 Arpajon, France\\
\it $^b$Universit\'e Paris-Saclay, CEA, Laboratoire Mati\`ere sous Conditions Extr\^emes,\\
\it F-91680 Bruy\`eres-le-Ch\^atel, France
\end{center}

\vspace{0.5cm}

\begin{abstract}
In this article, we examine different approaches for calculating low frequency opacities in the warm dense matter regime. The relevance of the average-atom approximation and of different models for calculating opacities, such as the Ziman or Ziman-Evans models is discussed and the results compared to \textit{ab initio} simulations. We begin by recalling the derivation of the Ziman-Evans resistivity from Kubo's linear response theory, using the local approximation to the solutions of the Lippmann-Schwinger equation. With the help of this approximation, we explicitly introduce an ionic structure factor into the Ziman formula, without resorting to the Born approximation. Both approaches involve the calculation of scattering phase shifts, which we integrate from Calogero equation with an adaptive step numerical scheme based on a Runge-Kutta-Merson solver. We show that if the atomic number $Z$ is not too large, integrating the phase shifts in this way is more time-efficient than using a classical Numerov-type scheme to solve the radial Schr\"odinger equation. Various approximations are explored for phase shifts to further improve computation time. For the Born approximation, we show that using Born phase shifts directly in the scattering cross-section gives more accurate results than with the integral formula based on the Fourier transform of the electron-ion potential. We also compare an analytical formula based on a Yukawa fit of the electron-ion potential to a numerical integration. The average-atom results are compared with DFT-based molecular dynamics simulations for aluminum in the dilute regime and for copper, aluminum and gold at solid density and different temperatures.
\end{abstract}

\section{Introduction}\label{sec1}

For the description of high energy density phenomena, such as in planetary interiors, stellar interiors or inertial confinement fusion capsules, we often rely on radiative-hydrodynamics simulations. To accurately model these extreme environments it is essential to provide the simulations with reliable transport coefficients, such as electrical conductivity, opacity, and thermal conductivity. For instance, in planetary interiors, the electrical conductivity plays an important role for the dynamo process\cite{Soubiran2018}. In stars, the opacity is a key component for the radiation transport\cite{Freytag1996} and in inertial-confinement-fusion (ICF) capsules thermal conductivity impacts the growth of hydrodynamic instabilities \cite{Atzeni2004} directly affecting the fusion efficiency. 

All these situations are relevant for warm dense matter (WDM), which corresponds to densities such that $0.1 \textrm{ g/cm}^3 \lesssim \rho \lesssim 10 \textrm{ g/cm}^3$ and temperatures such that $1 \textrm{ eV}\lesssim k_BT \lesssim 100 \textrm{ eV}$. In the WDM regime, theoretical challenges arise mainly because the average ionic kinetic energy $\langle E_{kin}\rangle$ is comparable with the average potential energy $\langle E_{pot} \rangle$, fixing the coupling parameter $\Gamma = \langle E_{pot}\rangle / \langle E_{kin} \rangle \approx 1$, making it unsuitable for perturbative expansions. A similar limitation applies to the electron degeneracy parameter, $k_BT/E_F$, where $k_BT$ is the thermal electronic energy and $E_F$ - the Fermi energy. These constraints hinder the applicability of traditional plasma physics methods and complicate their adaptation to the WDM regime. Experimental measurements are possible, but remain relatively rare due to the transient nature of WDM in laboratory settings. Alternatively, Kohn-Sham Density Functional Theory (KS-DFT) simulations offer a numerical approach to estimate transport properties, albeit at a high computational cost. To address this, the average-atom approximation \cite{Liberman1979} is often employed for more tractable computations. As demonstrated at the recent Charged-Particle Transport Code Comparison Workshop (CPTCCW) \cite{Stanek2024}, at local thermodynamic equilibrium (LTE), the average-atom approximation yields remarkably accurate results for the DC electrical conductivity in the WDM regime with significant computational efficiency. Such results could be readily applied for example to the interpretation of experiments involving optical diagnostics such as the streak optical pyrometry (SOP).

Concerning the low-frequency conductivity, as extensively discussed by Johnson and Kuchiev \cite{Johnson2009, Kuchiev2008}, the limitations of the average-atom approximation prevent the direct application of the Kubo-Greenwood formalism \cite{Kubo1957,Greenwood1958} commonly used in KS-DFT approaches. The issue arises from the divergence of radial matrix elements associated with free-free transitions. In the average-atom framework, interactions between electrons and neighboring ions outside a given potential well are neglected, leading to an unphysical prediction of infinite DC conductivity. To solve this problem, various methods have been developed~\cite{Johnson2009,Kuchiev2008,Evans1973,Ziman1960}. For example Johnson \textit{et al.} proposed a modified Kubo-Greenwood formula, by introducing a finite relaxation time directly in the conductivity. Later, Starrett~\cite{Starrett2016} derived an alternative formulation of the modified Kubo-Greenwood formula, incorporating an ionic structure factor, to account for ion correlations. Despite these advancements, the two most widely used approaches remain the Ziman formalism~\cite{Ziman1960}, based on the Boltzmann transport equation and the so-called Ziman-Evans approach~\cite{Evans1973}, which derives from the linear response theory. In the latter, the DC conductivity is generally calculated first and then extended to finite frequencies using a Drude model. Both approaches allow to correct for the divergence of the DC conductivity. However both models lack several important features, which limit their applicability.

The Ziman model introduces a finite electron-ion collision time, ensuring a finite static conductivity. However, although allowing for the derivation of the dynamic conductivity directly, without using a Drude model, the standard Ziman formalism lacks a natural incorporation of a structure factor, as does the Ziman-Evans model. This restricts the applicability of the formula in scenarios where ionic structure effects are significant. While including an ionic structure factor in Ziman's framework could be achieved using the Born approximation for the scattering amplitude, this is also a limitation for the applicability of the formula. As demonstrated in this work, the standard Born approximation based on the Fourier transform of the electron-ion potential performs poorly, at least in the WDM regime, necessitating refinements. Another approach, could be for example that used by Burrill \textit{et al.}~\cite{Burrill2016}, where the ionic structure factor was introduced directly in the relaxation time, multiplying the differential cross section as if the Born approximation was used, but a proper justification, as presented in this paper, could be interesting for confirming this formula.

The Ziman-Evans model is based on the local approximation introduced by Evans \textit{et al.} which is also derived in this article. It expresses the inverse of the DC conductivity, which is given by a force-force correlation integral. Although this model is based on Kubo's linear response theory, more general than Boltzmann's equation framework, which has a limited applicability range~\cite{Greenwood1958}, it only yields the DC conductivity. Numerous authors~\cite{Faussurier2019,Wetta2019,Wetta2023,Johnson2009,Johnson2020} then rely on a Drude model to obtain the dynamic conductivity. While a Drude model could work for simple metals like aluminum, it is known~\cite{Blanchet2024} that for transition metals, it could not work in principle, due to the localized electrons on the $d$-shell.

However, in our case, the average-atom approximation is notoriously known to lack ingredients for the description of transition metals, especially at low temperatures such as $k_BT<10$ eV. The question arises then: at a given spherically symmetric electron-ion potential $V(r)$, chemical potential $\mu$ and mean ionization $Z^{*}$, computed by an average-atom model, what is the simplest and most reliable framework to compute low frequency opacities? The main application could be the computation of opacity tables, to be supplied to radiative-hydrodynamic codes.

Throughout this work we compare the two cited models, after incorporating an ionic structure factor in the Ziman model, using the local approximation by Evans. We first compare the models in their validity range (dilute system and thus weak scattering) with \textit{ab initio} calculations performed with ABINIT~\cite{Gonze2020} for aluminum at $\rho_0/10$ and $k_BT$=5 eV, 10 eV and 15 eV. Then, we test the predictions for WDM against \textit{ab initio} simulations for $\textrm{Al}$ at $\rho_0$ and $k_BT$=1 eV, $\textrm{Au}$ at $\rho_0$ and and $k_BT=10$ eV and for $\textrm{Cu}$ at $\rho_0$ and $k_BT$=1 eV, 5 eV and 10 eV. This is a very constraining case for average-atom models since for transition metals, the precise nature of the electrons on the $d$-shell depend on the chosen boundary condition for the wavefunction \cite{Nikiforov2005}.

We proceed as follows, in the Section \ref{sec2}, we first define the context of the study in \ref{subsec21}, after which we outline, in \ref{subsec22} a simple derivation of the DC resistivity $R$ as given in Evans \textit{et al.}. Further, in \ref{subsec23}, we explain how to derive the conductivity, by using the same local approximation as in the Ziman‐Evans derivation, but based on a semi‐classical Boltzmann equation. We also show how the same approximation could be used to include an ionic structure factor in the frequency dependent complex conductivity $\sigma(\omega)$ and then, in subsection \ref{subsec24}, derive an expression for the opacity using the standard approach of Kramers-Kronig formulas. In both cases, we use the same formula for the principal part, the Drude model and the Ziman's formula being similar, as a function of the frequency $\omega$.

In Section \ref{sec3}, we show how to compute the scattering phase shifts $\delta_\ell$, which are needed in the two models, with an adaptive step numerical scheme, used to solve the Calogero equation. In subsection \ref{subsec31}, we present the numerical scheme and then we compare it with the classical Numerov scheme \cite{Numerov1924,Numerov1927} in terms of performance and accuracy in subsection \ref{subsec32}. We then show, in subsection \ref{subsec33}, how the integration of the opacity could be simplified, if instead of numerical phase shifts, one used Born phase shifts, with a Yukawa fit for the electron-ion potential. 

In Section \ref{sec4}, we compare the conductivities and the opacities computed with the models of Ziman-Evans and Ziman with \textit{ab initio} simulations. In \ref{subsec41}, we explain how the ABINIT code was used both to perform DFT based MD simulations and compute transport properties such as conductivities and opacities. In \ref{subsec42} we explain our average-atom computations. In subsection \ref{subsec43}, we compare individually the Ziman and Ziman-Evans' models with and without a structure factor. We used the hard spheres structure factor as a first step of comparison. More evolved structure factors such as the Bretonnet-Derouiche one \cite{Bretonnet1988} could be used in future works to better compare the two models in cases where the structure factor is essential. In subsection \ref{subsec44}, we examine the sum rule and the impact of the constraint of having to verify it, for both Ziman and Ziman-Evans models. We conclude in section \ref{sec5}.

\section{The Ziman and Ziman-Evans models for opacity calculations}\label{sec2}
\subsection{Context of the study}\label{subsec21}

In disordered systems, the theory of electrical conductivity could be either rather simple, if, for example, weak scattering is assumed \cite{Kohn1957}, or Boltzmann's equation used~\cite{Ziman1960} (which could be easily justified in the weak scattering limit), or very complicated, if one tries to go beyond weak scattering. Although for liquid metals, formal approaches for the conductivity were available\cite{Kubo1965}, it was not until the Tokyo conference of September 1972, that a simple idea to go beyond weak scattering was proposed by Evans \textit{et al.}\cite{Evans1973} with the local approximation, and Kubo's force-force formula could be evaluated, although only for the resistivity. However, even if not explicitly present in Evans', derivation, through the local approximation, the assumption of weak scattering is present through the weak scattering density, as we recall in Appendix~\ref{appA}. The same holds for the collision operator used in the Ziman formulation~\cite{Mahan2000}. The attempt to go beyond weak scattering is given at the end of the paper by Evans et al., in the nonlocal approximation.

Further in this section, we show how both Ziman-Evans and Ziman formalisms can be derived from Evans' local approximation of the self-consistent solution of the Lippmann-Schwinger equation. The considered system is a collection of $N_e$ non-interacting electrons, that are scattered by $N_s$ centers, each described by a finite-range, spherically symmetric potential $V_0(|\textbf{r}-\textbf{r}_i|)$, where $\textbf{r}_i$ is the position of the $i$-th scattering center. With $r_{\mathrm{ws}}$, being the potential range (or the Wigner-Seitz radius), we have $V_0(r) = 0$, for $r\ge r_{\mathrm{ws}}$.

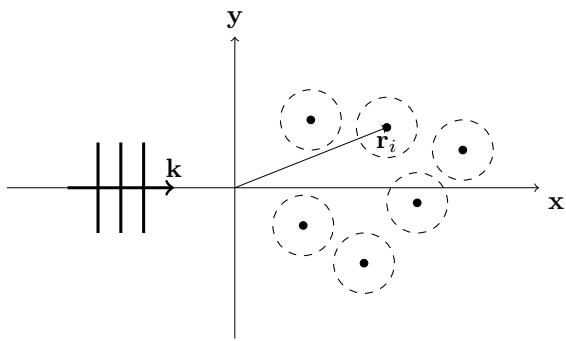
\begin{figure}[!ht]
    \centering
    \begin{tikzpicture}[scale=1]
    \draw[line width=0.1mm,->] (-3,0) -- (4,0) node[below right] {$\textbf{x}$};
    \draw[line width=0.1mm,->] (0,-2) -- (0,2) node[above] {$\textbf{y}$};
    \draw[line width=0.4mm,->] (-2.2,0) -- (-0.8,0) node[above] {$\textbf{$\textbf{k}$}$};
    \draw[line width=0.1mm,->] (0,0) -- (2,0.8) node[below] {$\textbf{$\textbf{r}_i$}$};
    
    \draw[line width=0.4mm,-] (-1.2,-0.6) -- (-1.2,0.6);
    \draw[line width=0.4mm,-] (-1.5,-0.6) -- (-1.5,0.6);
    \draw[line width=0.4mm,-] (-1.8,-0.6) -- (-1.8,0.6);
    
    \draw[dashed] (1,0.9) circle (0.4) ;
    \draw[fill=black] (1,0.9) circle (0.05) ;
    \draw[dashed] (0.9,-0.5) circle (0.4) ;
    \draw[fill=black] (0.9,-0.5) circle (0.05) ;
    \draw[dashed] (2,0.8) circle (0.4) ;
    \draw[fill=black] (2,0.8) circle (0.05) ;
    \draw[dashed] (1.7,-1) circle (0.4) ;
    \draw[fill=black] (1.7,-1) circle (0.05) ;
    \draw[dashed] (2.4,-0.2) circle (0.4) ;
    \draw[fill=black] (2.4,-0.2) circle (0.05) ;
    \draw[dashed] (3,0.5) circle (0.4) ;
    \draw[fill=black] (3,0.5) circle (0.05) ;
    
    \end{tikzpicture}
    \caption{A schematic of the diffusion process. A plane wave representing a set of non interacting electrons, encounters an array of fixed scatterers, each with a finite range potential, with the frontier represented by the dashed lines. }
    \label{fig:schema_diffusion}
\end{figure}

Since there is no interaction between the electrons, to describe our system we only need to consider a single electron, of incident wave vector, say $\textbf{k}$, passing through an array of scatterers. This situation is represented on Fig.\ref{fig:schema_diffusion}. In atomic units, the Hamiltonian describing it, can be written as,
\begin{equation}
\hat{H}_0 = \frac{\hat{\textbf{p}}^2}{2} + \sum_{i=1}^{N_s} \hat{V}_0(|\textbf{r} - \textbf{r}_i|),
\label{eq:hamiltonian}
\end{equation}
where $\hat{\textbf{p}}$ is the momentum operator, and $\hat{V}_0$ is the potential energy operator describing a single scatterer. The Schr\"odinger equation, governing the evolution of the electron's wave function writes in the stationary form as,
\begin{equation}
\left[ \frac{\hat{\textbf{p}}^2}{2} + \sum_{i=1}^{N_s} \hat{V}_0(| \textbf{r} - \textbf{r}_i |) \right] \psi(\textbf{r}) = E \psi(\textbf{r}),
\end{equation}
where \( \psi(\textbf{r}) \) is the electron's wave function, and \( E \) is its energy. The Lippmann-Schwinger equation \cite{Marchildon2002} provides a solution to the Schr\"odinger equation, written in a self-consistent way and with the proper boundary conditions corresponding to a scattering interaction. The solution writes as a combination of the incident plane wave and the scattered waves from the potential, as follows, 
\begin{equation}\label{eq:lippmann2}
\psi(\textbf{r}) = e^{i\textbf{k}\cdot\textbf{r}} - \frac{1}{2\pi} \int_{\mathbb{R}^3} \frac{e^{ik|\textbf{r}-\textbf{r}_0|}}{|\textbf{r}-\textbf{r}_0|}\sum_{i=1}^{N_s}V_0(|\textbf{r}_0-\textbf{r}_i |)\psi(\textbf{r}_0)\mathrm{d}^3r_0 
\end{equation}
where $e^{i\textbf{k}\cdot\textbf{r}}$ represents the incident plane wave. Here, the positive, outwards solution of the Lippmann-Schwinger equation is retained.

For $N_s = 1$, and in the limit of large $|\textbf{r}|$, it is possible to solve this equation exactly, by simply developing both the plane wave and the wave function with the help of Legendre polynomials \cite{Abramowitz1964} $P_\ell$ (in a spherically symmetrical potential). This is discussed in more details in Appendix~\ref{appB}. In this case, the wave function writes 
\begin{equation}
\psi(\textbf{r}) = e^{i\textbf{k}\cdot\textbf{r}} + \sum_{\ell=0}^\infty i^{\ell} e^{i\delta_\ell} (2\ell + 1) \frac{u_\ell(r)}{r} P_\ell(\cos\theta),
\end{equation}
where the $u_\ell$ are solutions to an integral equation coming from Eq.~(\ref{eq:lippmann2}), or to the radial Schr\"odinger equation. The $\delta_\ell$ are the scattering phase shifts, and $\theta$ is the angle variable in the spherical coordinate system.

For $N_s > 1$, the solution of Eq.~(\ref{eq:lippmann2}) becomes significantly more complicated. The authors are not aware of any analytical solutions in this case, as the presence of multiple scattering centers introduces interferences and correlations that must be treated numerically or through different approximations. However, if one neglects multiple scattering, the total wavefunction can simply be written as a superposition of individually scattered waves, 
\begin{equation}
\psi(\textbf{r}) = e^{i\textbf{k}\cdot\textbf{r}} + \sum_{i=1}^{N_s} \psi_i(\textbf{r}-\textbf{r}_i),
\end{equation}
where each $\psi_i(\textbf{r}-\textbf{r}_i)$, is the wave-function scattered from the center located at $\textbf{r}_i$. This is an approximation one could start with. The problem in this case is the well-known divergence of the DC conductivity~\cite{Johnson2006,Johnson2009,Kuchiev2008}.

Alternatively, the approximation first introduced by Evans et al., which assumes locality of the wave function solution of Eq.~(\ref{eq:lippmann2}) allows us, in principle, to avoid this divergence. The difference between this approximation and that of single scattering is that we only approximate the wave function inside each potential well. By doing so, we allow for multiple scattering outside these wells, which prevents the divergence of the DC conductivity, and even allows one for introducing a relaxation time as shown by Edwards \cite{Edwards1965}. In this approximation, the wavefunction writes around each potential well as, 
\begin{equation}\label{eq:approx_ls}
\psi(\textbf{r}) \underset{|\textbf{r}-\textbf{r}_n| < r_{\mathrm{ws}}}{=} e^{i\textbf{k}\cdot\textbf{r}_n}\sum_{\ell=0}^\infty i^{\ell} e^{i\delta_\ell} (2\ell + 1) \frac{u_\ell(| \textbf{r}-\textbf{r}_n|)}{|\textbf{r}-\textbf{r}_n|} P_\ell(\cos\theta).
\end{equation}
Here $\theta$ is the angle between the local coordinate inside each potential well and the $\textbf{x}$ axis. The factor $e^{i\textbf{k}\cdot\textbf{r}_n}$ arises from the well being centered at $\textbf{r}_n$ instead of at the origin, as explained in Appendix~\ref{appB}.
Like Evans, we make no assumptions about the value of the wavefunction if $|\textbf{r}-\textbf{r}_i| > r_{\mathrm{ws}}$. The wavefunction should be the superposition of all the scattering events between the freely propagating (in the absence of the potential $V_0$) but phase shifted waves.

\subsection{The Ziman-Evans approach from the linear response theory}\label{subsec22}
The Ziman-Evans approach is based on the evaluation of an inverse transport coefficient, which is the electrical resistivity $R(\omega)$, or more precisely - its DC limit, $R(0)$. Since the resistivity is the inverse of the conductivity $\sigma(\omega)$, a question arises as to why we simply don't use Kubo's formula~\cite{Kubo1957,Band_Avishai_2013}, which is given by (see Appendix \ref{appA}), 
\begin{equation}\label{eq:kubo1}
\sigma(\omega) = i\int_{0}^{+\infty} \mathrm{d}t e^{i\omega t} \langle [\hat{p}_x(t),\hat{x}(0)]\rangle,
\end{equation}
where $\hat{x}$ is the electron coordinate operator and $\hat{p}_x$ is the momentum operator along the $\textbf{x}$ axis. The average of an operator $\langle\hat{B}\rangle$ is $\textrm{Tr}(\hat{\rho}_0\hat{B})$, where $\hat{\rho}_0$ is the density matrix corresponding to the system at equilibrium, not perturbed by an external electric field. The commutator of $\hat{A}$ and $\hat{B}$ is given by $[\hat{A},\hat{B}] = \hat{A}\hat{B} - \hat{B}\hat{A}$.

In this expression one can safely take $\omega = 0$ and find a finite DC conductivity and thus resistivity. However, as discussed in the previous subsection, this is only true while there is an interaction between scatterers. In the considered case, which to be consistent with the local approximation, should correspond to a dilute system, the density of scatterers $n_s$ is small. In the limit where $n_s \to 0$ (or equivalently in the limit of weak scattering), the DC conductivity $\sigma(0)$ diverges, due to the fact that the system tends towards isolated scatterers. This is the reason why, one would rather work with the resistivity $R = 1/\sigma$, such that $R \underset{ n_s \to 0}{\to} 0$. To derive this resistivity~\cite{Sorbello1981} one uses Eq.~(\ref{eq:kubo1}), by expanding it at the first order in $n_s$, after integration by parts and some manipulations that are detailed in Appendix \ref{appA},
\begin{equation*}
R = \int_{0}^{+\infty} \mathrm{d}t \int_{0}^{\beta} \mathrm{d}\lambda \langle \hat{F}_x (0) \hat{F}_x(t+i\lambda) \rangle.
\end{equation*}
Here, $\hat{F}_x$ is the force operator in the $\textbf{x}$ direction, or $\hat{F}_x = d\hat{p}_x/dt = i[\hat{H}_0,\hat{p}_x]$. The resistivity is thus expressed with a force-force correlation function \cite{Edwards1965,Evans1973,Ballentine1974,Szabo1972,Szabo1973,Szabo1976,Szabo1978,Rousseau1972,Chambers1973}, which has yet to be evaluated in the local approximation given previously. We outline this derivation in Appendix \ref{appA}, to obtain as Evans \textit{et al.}, the following formula for $R$, also valid at a finite temperature, 
\begin{equation*}
\begin{aligned}
R = &- \frac{2\pi}{3{Z^{*}}^2n_i}\int_{0}^{+\infty} \mathrm{d}~\varepsilon f_{FD}'(\varepsilon)\int_{\mathbb{R}^3}\mathrm{d}^3k\int_{\mathbb{R}^3}\mathrm{d}^3{k'} \\
&\times\left\langle \sum_{i,j=1}^{N_s}\langle\psi_k|\nabla V(\textbf{r}-\textbf{r}_i)|\psi_{k'}\rangle \langle\psi_k|\nabla V(\textbf{r}-\textbf{r}_j)|\psi_{k'}\rangle \right\rangle_0 \delta(\varepsilon-k^2)\delta(\varepsilon-k'^2),
\end{aligned}
\end{equation*}
where $\langle \rangle_0$ denotes averaging over the position of the scatterers, $n_i$ is the ion density, $Z^*$ the mean ionic charge (or mean ionization),
\begin{equation*}
    f_{FD}(\varepsilon) =\frac{1}{1+e^{\beta(\varepsilon-\mu)}}
\end{equation*}
is the Fermi-Dirac distribution ($\beta=1/(k_BT)$ and $\mu$ represents the chemical potential). The derivative of the Fermi-Dirac distribution (which was replaced by a Dirac at the Fermi energy in the original derivation by Evans) reads, 
\begin{equation*}
    f'_{FD}(\varepsilon) =\dfrac{\partial f_{\mathrm{FD}}}{\partial \varepsilon}(\varepsilon)=-\beta f_{\mathrm{FD}}(\varepsilon)\left[1-f_{\mathrm{FD}}(\varepsilon)\right]. 
\end{equation*}
By plugging the approximation given by Eq.~(\ref{eq:approx_ls}), we can derive the usual Ziman-Evans formula for the DC resistivity,
\begin{equation}\label{eta}
    R=-\dfrac{1}{3\pi {Z^*}^2 n_i} \int_0^\infty f'_{\mathrm{FD}}(\varepsilon) \mathcal{I}(\varepsilon) \mathrm{d}\varepsilon.
\end{equation}
The function $\mathcal{I}(\varepsilon)$ is given by\cite{Wetta2019,Wetta2023} 
\begin{equation*}
    \mathcal{I}(\varepsilon)=\int_0^{2k}q^3 S(q) \sigma(q) \mathrm{d}q,
\end{equation*}
where $S(q)$ denotes the static ion-ion structure factor and $\sigma(q)$ the electron-ion scattering cross-section. The vector $\textbf{q}=\textbf{k}^\prime-\textbf{k}$ is the momentum transferred in the elastic scattering event (\emph{i.e.} such as $|\textbf{k}^\prime|=|\textbf{k}|$) of a conduction electron from an initial state $\textbf{k}$ to a final $\textbf{k}^\prime$ one. Introducing the scattering angle $\theta\equiv (\textbf{k},\textbf{k}^\prime)$ and its cosine $\chi=\cos\theta$, one has $q^2=2k^2 (1-\chi)$ and
\begin{equation*}
    \mathcal{I}(\varepsilon)=2k^4 \int_{-1}^1 S\left[k\sqrt{2(1-\chi)}\right]|\mathcal{A}(k,\chi)|^2 (1-\chi) \mathrm{d}\chi.
\end{equation*}
The $T-$matrix formalism of Evans \cite{Evans1973} provides the modulus of the electron-ion scattering amplitude $|\mathcal{A}(k,\chi)|$, whose square is actually $\sigma(q)$ \cite{Pain2010}:
\begin{equation*}
\sigma(q)=|\mathcal{A}(k,\chi)|^2=\frac{1}{k^2}\Big|\sum_\ell (2\ell+1)e^{i\delta_\ell}\sin(\delta_\ell)P_{\ell}(\chi)\Big|^2.
\end{equation*}
\subsection{The Ziman formalism with an ionic structure factor}\label{subsec23}
In this subsection, we show how to derive the conductivity, by using the same local approximation as in the Ziman-Evans derivation, but based on a semi-classical Boltzmann equation. As already pointed by~\cite{Kubo1957,Greenwood1958}, the Boltzmann equation was only derived in the weak scattering limit, and under the condition that $\hbar/\tau < \eta$, where $\tau$ is the collision time and $\eta$ is a typical energy scale, such as the thermal energy or the Fermi energy. In our case, this condition is equivalent to saying that electrons should not be scattered too often compared to the mean thermal fluctuation time, in order not to violate the assumption of molecular chaos, or of particles being independent before collision. Another observation is that $\tau$ is the mean time after which, our perturbed system returns to equilibrium and $\hbar/k_BT$ is the quantum coherence scale. The stated condition ensures that the system remains in a regime where quantum coherence effects between collisions are negligible, allowing us to treat each collision as an independent event. This is crucial for maintaining the validity of the semiclassical Boltzmann equation, which assumes that the phase space distribution function evolves under the influence of external fields and collisions without significant quantum interference between successive events.

In atomic units, the Boltzmann equation can be written for the electron distribution function $f(\textbf{r},\textbf{p},t)$ as follows, 
\begin{equation*}
\frac{\partial f}{\partial t} + \textbf{p}\cdot\frac{\partial f}{\partial \textbf{r}} - \textbf{E} \cdot \frac{\partial f}{\partial \textbf{p}} = -\frac{df}{dt}\Big|_{\textrm{collisions}},
\end{equation*}
where the collision operator is written in the usual RTA (relaxation time approximation), or $-(f-f_0)/\tau_p$, where $f_0$ is the Fermi-Dirac distribution function and $\tau_p$ is the moment dependent relaxation time. If we also consider the dipolar approximation, where the field's typical wavelength is greater than individual scatterers dimensions, then we can neglect the $\textbf{r}$ dependence. We then linearize the equation in the presence of an external electric field $\textbf{E} = e^{-i\omega t}\textbf{E}_0$, with $f = f_0 + f_1$,
\begin{equation*}
\frac{\partial f_1}{\partial t} + \frac{f_1}{\tau_p} = \frac{\partial f_0}{\partial \textbf{p}}\cdot\textbf{E}_0e^{-i\omega t},
\end{equation*}
where, the unperturbed distribution is given by $f_0(p) = f_{FD}(\varepsilon)$ (with $\varepsilon = p^2/2$). The linearized equation can be solved exactly and one can check that the following expression is a solution, with $f_1(\textbf{p},t=0) = 0$, 
\begin{equation*}
f_1(\textbf{p},t)  = \frac{\tau_pf_0(p)}{1-i\omega\tau_p} \frac{\partial f_0}{\partial \textbf{p}}\cdot\textbf{E}_0 e^{-i\omega t}.
\end{equation*}
If we now compute the transverse current, with say $\textbf{E}_0 = E_0\textbf{y}$, and $\textbf{j} =  j\textbf{y}$, then with $j = -2/(2\pi)^3\int_{\mathbb{R}^3} pf_1(\textbf{p},t)\mathrm{d}^3p$, we directly get the frequency-dependent conductivity, after averaging over the different directions,
\begin{equation*}
\sigma(\omega) =\frac{\beta}{3\pi^2}\int_{0}^{+\infty} \frac{p^{4}\tau_pf_0(p)(1-f_0(p))}{1-i\omega \tau_p} \mathrm{d}p.
\end{equation*}
The collision time is given by (see \cite{Peierls1955,Mahan2000}):
\begin{equation*}
\frac{1}{\tau_p} = n_i\int_{\mathbb{R}^3}\mathrm{d}^3k'\delta(\varepsilon_k-\varepsilon_{k'})|T_{kk'}|^2(1-\cos(\theta')),
\end{equation*}
where $\theta'$ is the angle between $\textbf{k}'$ and $\textbf{k}$ and $T_{kk'}$ is the scattering amplitude, given by the Lippmann-Schwinger equation. In the limit where $|\textbf{r}|$ is large, $e^{ik|\textbf{r}-\textbf{r}_0|}/|\textbf{r}-\textbf{r}_0| \approx e^{ikr}e^{-ik\textbf{r}\cdot\textbf{r}_0/r}/r$. The diffusion amplitude can then be directly expressed from the following, 
\begin{equation*}
\psi(\textbf{r}) \underset{ |\textbf{r} |\to \infty }{=} e^{i\textbf{k}\cdot\textbf{r}} - \frac{1}{2\pi} \frac{e^{ikr}}{r}\int_{\mathbb{R}^3} e^{-ik\textbf{r}\cdot\textbf{r}_0/r} \sum_{i=1}^{N_s}V_0(| \textbf{r}_0-\textbf{r}_i |)\psi(\textbf{r}_0)\mathrm{d}^3r_0 .
\end{equation*}
The scattering amplitude writes then, as shown in Appendix \ref{appB}:
\begin{equation*}
T_{kk'} = -\frac{1}{2\pi}\int_{\mathbb{R}^3} e^{-i\textbf{k}'\cdot\textbf{r}} \sum_{n=1}^{N_s}V_0(| \textbf{r}_0-\textbf{r}_n |)\psi(\textbf{r}_0)\mathrm{d}^3r_0,
\end{equation*}
where (in the case of elastic scattering) with $|\textbf{k}| = |\textbf{k}'|$, $\textbf{k}' = k\textbf{r}/r$ is wave vector pointing in the scattered direction. With a variable change $\textbf{r}_1 = \textbf{r}_0 - \textbf{r}_n$ in each term of the sum, one obtains,
\begin{equation*}
T_{kk'} = -\frac{1}{2\pi}\sum_{n=1}^{N_s}e^{-i\textbf{k}'\cdot\textbf{r}_n}\int_{\mathbb{R}^3} e^{-i\textbf{k}'\cdot\textbf{r}_1} V_0(| \textbf{r}_1|)\psi(\textbf{r}_n+\textbf{r}_1)\mathrm{d}^3r_1.
\end{equation*}
Using the local approximation of Eq.~(\ref{eq:approx_ls}), which was also derived in the Appendix~\ref{appB}, we readily find, 
\begin{equation*}
T_{kk'} = -\frac{S}{2\pi}\int_{\mathbb{R}^3} e^{-i\textbf{k}'\cdot\textbf{r}_1} V_0(| \textbf{r}_1|) \sum_{\ell=0}^{+\infty}(2\ell+1)i^{\ell}\frac{u_\ell(|\textbf{r}_1|)}{|\textbf{r}_1|}P_\ell(\cos(\theta_1)) \mathrm{d}^3 r_1,
\end{equation*}
where $S$ is the ionic structure factor that is given by (after averaging over the randomly distributed scattering centers),
\begin{equation*}
S(\textbf{k}-\textbf{k}')= \frac{1}{N_s}\left|\sum_{n=1}^{N_s}e^{-i(\textbf{k}-\textbf{k}')\cdot\textbf{r}_n}\right|^2.
\end{equation*}
The term that multiplies the structure factor is nothing else than the single scatterer diffusion amplitude, for a scattering center at the origin, which is known (see Appendix~\ref{appB}). With $\delta(\varepsilon_k - \varepsilon_{k'}) = 2/(k+k')\delta(k-k')$, we can end up expressing the relaxation time $\tau_p$ as being, 
\begin{equation*}
\frac{1}{\tau_p} = n_i\int_{0}^{\pi}\frac{1}{k}2\pi k^2\sin(\theta)S(q) |f(\theta)|^2(1-\cos(\theta))\mathrm{d}\theta,
\end{equation*}
where $q^2 = 2k^2[1-\cos(\theta)]$ and the diffusion amplitude is given by $f(\theta) =(1/k) \sum_{\ell=0}^{+\infty}(2\ell+1)e^{i\delta_\ell}\sin(\delta_\ell)P_\ell(\cos(\theta))$.
With a variable change, the relaxation time can be cast in its final form, which is, 
\begin{equation*}
\frac{1}{\tau_p} = \frac{\pi n_i}{k^3}\int_{0}^{2k} q^3S(q)|f(q)|^2\mathrm{d}q,
\end{equation*}
where $f(q)$ is given by, 
\begin{equation*}
f(q) =\frac{1}{k}\sum_{\ell=0}^{+\infty}(2\ell+1)e^{i\delta_\ell}\sin(\delta_\ell)P_\ell\left(1-\frac{q^2}{2k^2}\right).
\end{equation*}
A classical result can be obtained for a unity structure factor for the relaxation time, as explained in Appendix~\ref{appB}, 
\begin{equation*}
\frac{1}{\tau_p} = p\sigma_{tr}n_i = \frac{4\pi n_i}{p} \sum_{\ell=0}^{+\infty} (\ell+1)\sin^2[\delta_{\ell+1}(p)-\delta_\ell(p)],
\end{equation*}
where $\sigma_{tr}$ is the so-called transport cross-section, see Appendix~\ref{appC}.

\subsection{The frequency dependent opacity using Kramers-Kronig formulas}\label{subsec24}

Using Kramers-Kronig formulas, it is possible to get the full complex conductivity. In the previous subsection, we have it already due to the considered complex electric field, but for example in the Ziman-Evans approach, we only compute the real conductivity first, with the Drude model, and then the imaginary part, as  explained in Appendix.~\ref{appD}. In particular, for the imaginary part of the conductivity, we have, 
\begin{equation}\label{imag}
\Im\left[\sigma(\omega)\right]=-\frac{2\omega}{\pi}\mathcal {P}\!\!\int _{0}^{\infty}\frac{\Re\left[\sigma(\omega')\right]}{\omega'^{2}-\omega ^{2}}\,d\omega ',
\end{equation} 
where $\mathcal {P}$ denotes the principal value.

For the frequency-dependent (or spectral) opacity (photo-absorption cross-section per mass unit) we have (see for instance Ref. \cite{Callaway1974,Johnson2009}):
\begin{equation}
    \kappa(\omega)=\frac{1}{\rho}\frac{4\pi}{n(\omega)c}\Re\left[\sigma(\omega)\right],
\end{equation}
where the refraction index $n(\omega)$ reads
\begin{equation*}
    n(\omega)=\sqrt{\frac{|\epsilon(\omega)|+\Re \left[\epsilon(\omega)\right]}{2}},
\end{equation*}
$\epsilon(\omega)$ being the frequency-dependent dielectric function,
\begin{equation*}
 \epsilon(\omega)=1+4\pi\,i\frac{\sigma(\omega)}{\omega}.
\end{equation*}
\section{Numerical integration of the models}\label{sec3}
\subsection{Solving the Calogero equation using an adaptive scheme}\label{subsec31}
Both Ziman and Ziman-Evans models use scattering phase shifts $\delta_\ell$ as their main ingredients. Such quantities can be obtained from solving the Schr\"odinger equation for the radial part of the wavefunction $u_\ell$. It reads,
\begin{equation}\label{schro}
u_\ell''+ \left[p^2 - 2V(r) - \frac{\ell(\ell+1)}{r^2}\right]u_\ell = 0.
\end{equation}
The numerical solution to this equation, with the proper boundary conditions, can by obtained by a Numerov type scheme, and then compared to the Bessel functions~\cite{Abramowitz1964}, $j_\ell$ and $n_\ell$, which are two independent solutions of Eq.~(\ref{schro}) for $V(r)=0$ (free particles). This method is almost universally used to obtain the scattering phase shifts $\delta_\ell$. However, by solving the above equation, we get an additional information that is not used. Namely, we also compute the amplitude function $A_\ell(r)$. As Calogero showed, it is possible to derive an equation only for the phase shifts. Let us set \cite{Babikov1967}:
\begin{equation}\label{wf}
    u_\ell(r)=A_\ell(r)\left[\cos(\delta_{\ell}(r))j_{\ell}(pr)-\sin(\delta_{\ell}(r))n_{\ell}(pr)\right].
\end{equation}
The latter expression still does not allow us to determine both functions $A_\ell(r)$ and $\delta_\ell(r)$ uniquely. The idea consists in requiring also that the derivative of the wavefunction can be put in the form
\begin{equation}\label{C1}
    u_\ell'(r)=A_\ell(r)\left[\cos(\delta_{\ell}(r))j_{\ell}'(pr)-\sin(\delta_{\ell}(r))n_{\ell}'(pr)\right].
\end{equation}
This implies that $A_\ell(r)$ and $\delta_\ell(r)$ must satisfy the condition
\begin{equation}\label{C2}
\begin{gathered}
A_\ell'(r)\left[\cos(\delta_{\ell}(r))\,j_{\ell}(pr)-\sin(\delta_{\ell}(r))\,n_{\ell}(pr)\right] =\delta_\ell'(r)A_\ell(r)\left[\sin(\delta_{\ell}(r))\,j_{\ell}(pr)+\cos(\delta_{\ell}(r))\,n_{\ell}(pr)\right].
\end{gathered}
\end{equation}
Expression (\ref{C1}) and condition (\ref{C2}) are obvious for $r\rightarrow\infty$, if we expect that the functions $A_\ell(r)$ and $\delta_\ell(r)$ tend to constant values. Let us assume that the
potential has a finite range $r_{\mathrm{ws}}$, so that $V(r) = 0$ for
$r \ge r_{\mathrm{ws}}$. Then in that region the functions $A_\ell(r)$ and $\delta_\ell(r)$ must also take constant values $A_\ell(r_{\mathrm{ws}})$ and $\delta_\ell(r_{\mathrm{ws}})$. Thus, the relation (\ref{C1}) corresponds to the condition
of continuity of the derivative of the wave function at $r=r_{\mathrm{ws}}$. 
It follows from Eqs. (\ref{schro}), (\ref{wf}) and (\ref{C2}) that the functions $A_\ell(r)$ and $\delta_\ell(r)$ satisfy first-order equations, and the equation for $\delta_\ell(r)$ turns out to be independent of $A_\ell(r)$:
\begin{equation}\label{cal}
\delta_{\ell}'(r) = -\frac{2V(r)}{p}[\cos(\delta_{\ell})j_{\ell}(pr)-\sin(\delta_{\ell})n_{\ell}(pr)]^2,
\end{equation}
and is known as the Calogero equation\cite{Calogero1963,Calogero1964,Calogero1967}. This equation can be numerically solved using a classical RK4 scheme, but we found the convergence to be rather slow. This is why an adaptive scheme was used instead, based on the so-called Runge-Kutta-Merson (RKM) solver\cite{Merson1957,Scraton1964,Sarafyan1966,Chai1968}. The RKM method outlines a process for deciding the step size given a predetermined accuracy $tol$. For this method, (instead of four as in the RK4 scheme), five functions are evaluated at every step. It is a fifth-order method. Applied to Calogero equation, we can write,
\begin{equation*}
    \frac{d\delta_{\ell}}{dr}=f(r,\delta_{\ell}),
\end{equation*}
where we have set
\begin{equation*}
    f(r,\delta_{\ell})=-\frac{2V(r)}{p} \left[ \cos(\delta_\ell)j_\ell(pr)-\sin(\delta_\ell)n_\ell(pr)\right]^2.
\end{equation*}
Setting $\delta_{\ell,n}=\delta_{\ell}(r_n)$, and $r_n=r_0+nh$, the algorithm reads
\begin{align*}
    k_1=&h\,f(r_n,\delta_{\ell,n}),\nonumber\\
    k_2=&h\,f\left(r_n+\frac{h}{3},\delta_{\ell,n}+\frac{k_1}{3}\right),\nonumber\\
    k_3=&h\,f\left(r_n+\frac{h}{3},\delta_{\ell,n}+\frac{k_1}{6}+\frac{k_2}{6}\right),\nonumber\\
    k_4=&h\,f\left(r_n+\frac{h}{2},\delta_{\ell,n}+\frac{k_1}{8}+3\frac{k_3}{8}\right),\nonumber\\
    k_5=&h\,f\left(r_n+h,\delta_{\ell,n}+\frac{k_1}{2}-3\frac{k_3}{2}+2k_4\right)\nonumber\\
\end{align*}
and
\begin{equation*}
    \delta_{\ell,n+1}=\delta_{\ell,n}+\frac{1}{6}(k_1+4k_4+k_5)+O(h^5)
\end{equation*}
and the local error can be estimated from a weighted sum of the individual estimates:
\begin{equation*}
\mathscr{E}=\frac{1}{30}\left(2k_1-9k_3+8k_4-k_5\right).
\end{equation*}
Given a tolerance $\mathrm{tol}$, the adaptive step is modified during the iteration scheme according to
\begin{equation*}
\left\{
    \begin{array}{ll}
        h \rightarrow h/2 & \mbox{ if }\;\; |\delta_{\ell,n+1}-\delta_{\ell,n}| > \mathrm{tol}, \\
        h \rightarrow h & \mbox{ if }\;\; \mathrm{tol}/64 < |\delta_{\ell,n+1}-\delta_{\ell,n}| < \mathrm{tol}\\
        h \rightarrow 2h & \mbox{ if }\;\;|\delta_{\ell,n+1}-\delta_{\ell,n}| < \mathrm{tol}/64.
    \end{array}
\right.
\nonumber
\end{equation*}
\begin{figure}[!ht]
\centerline{\includegraphics[width=0.6\textwidth]{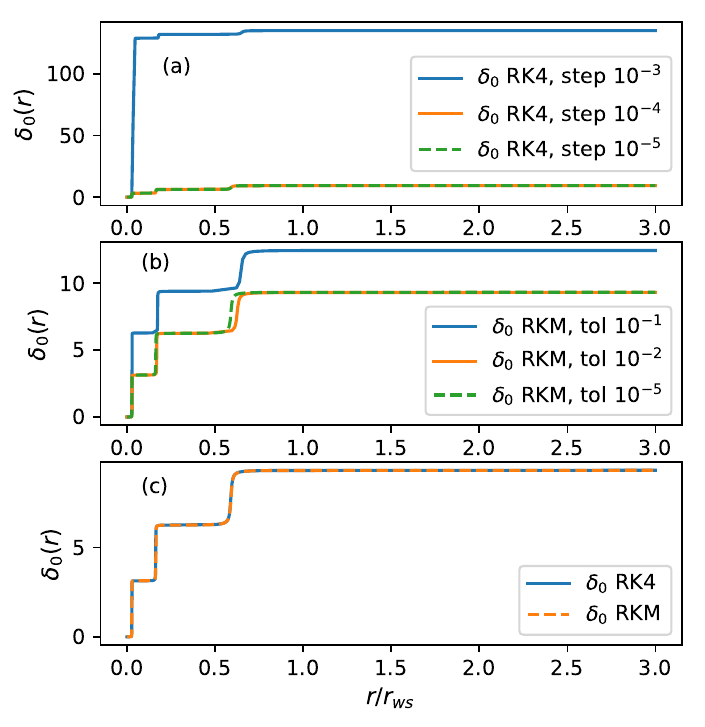}}
\caption{The scattering phase shift $\delta_0$ is represented in \textbf{(a)} computed with a RK4 scheme and different steps, in \textbf{(b)} with a RKM scheme and different tolerances, and in \textbf{(c)} we represent the comparison between the converged RK4 and RKM results. The potential used is $V(r) = -28e^{-3r}/r$. The gain in computational time is of the order of $10^3$ of RKM versus RK4.\label{fig:rkm_rk4}}
\end{figure}

The step is accepted and $r_n$ is incremented only when $|\delta_{\ell,n+1}-\delta_{\ell,n}|< \textrm{tol}$, which correspond to the second and third cases in the scheme above. A comparison between these two methods (the RK4 and RKM schemes) can be seen in Fig.~\ref{fig:rkm_rk4}. In Fig.~\ref{fig:rkm_rk4} \textbf{(a)}, we see that with a classical RK4 scheme, one need a step of $h=10^{-4}$ to achieve proper convergence, while with the RKM solver, a tolerance of $tol=10^{-2}$ is almost sufficient for convergence as seen in Fig.~\ref{fig:rkm_rk4} \textbf{(b)}. As a result, a gain in time of the order of a factor of 1000 is achieved between RKM and RK4 in this particular example. The reason for this, is the presence of two distinct scales in the non-linear Calogero equation, a rapid and a slow one. Due to the supplementary condition, when the error is smaller than $tol/64$, the RKM can double its step on each iteration on the slow time scale, allowing for a fast "hopping" of slowly varying regions. Also, when the function is rapidly varying, the step is exponentially reduced until a proper integration is achieved at the fixed tolerance.
\subsection{Comparison with the Numerov numerical scheme}\label{subsec32}
\begin{figure}
\centerline{\includegraphics[width=0.87\textwidth]{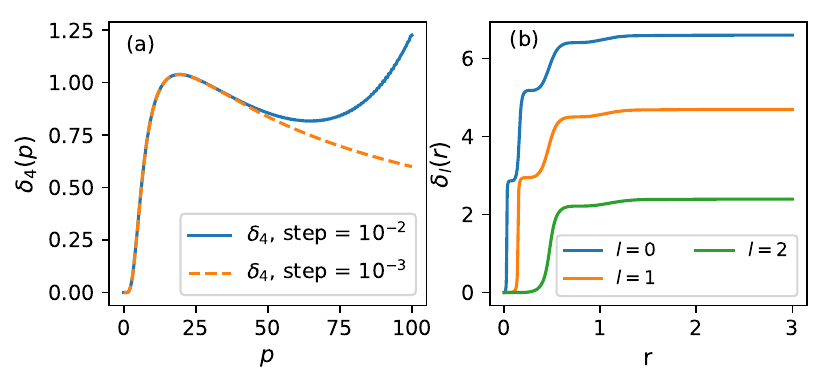}}
\caption{In \textbf{(a)} we represent the phase shift $\delta_4(p)$ integrated with a Numerov scheme from the Schr\"odinger equation with two different steps $h$. In \textbf{(b)} we see the phase shifts $\delta_0$, $\delta_1$ and $\delta_2$ as a function of $r$, integrated from the Calogero equation using a RKM solver with $\textrm{tol} = 10^{-7}$. In the two cases a Yukawa-type potential $V(r) = -28~e^{-3r}/r$ is used.\label{fig:numerov_rkm}}
\end{figure}

\begin{figure}
\centering{\includegraphics[width=0.8\textwidth]{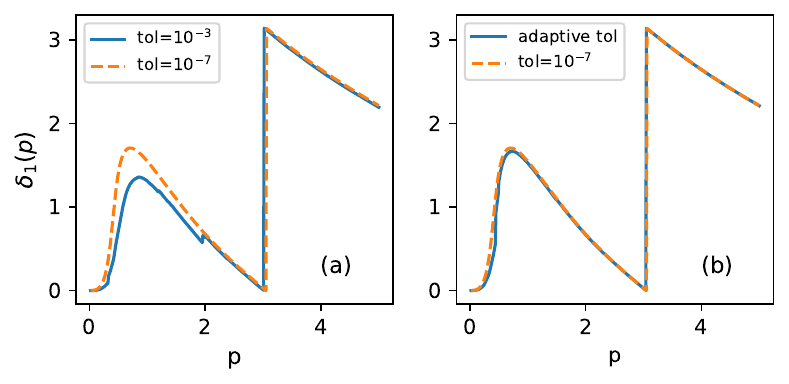}}
\caption{In \textbf{(a)} scattering phase shift $\delta_1$ as a function of $p$ as computed by a converged RKM scheme with $\mathrm{tol}=10^{-3}$ and with $\mathrm{tol}=10^{-7}$. We see the lack of precision in some regions. In \textbf{(b)} - the same comparison, but with a supplementary condition given by verifying Eq.~(\ref{eqcond}) to a precision of $5\times 10^{-2}$ with $tol=10^{-3}$, if not $tol=10^{-5}$ was used. The potential that was used to integrate these phase shifts was an average-atom potential corresponding to $\textrm{Cu}$ at solid density and $k_BT=1$ $eV$.\label{fig:sigma_dilue}}
\end{figure}

The conductivity $\sigma(\omega)$ is obtained after computing the phase shifts. For that, one can use a classical Numerov scheme, to solve the Schr\"odinger equation (\ref{schro}).
However, a well known weakness of the Numerov scheme is the constant step, which, if not set properly, can lead to errors, especially at high values of the momentum $p$, as can be seen on Fig.~\ref{fig:numerov_rkm} \textbf{(a)}. 
For a robust and rapid computation of phase shifts, without having to adapt the radial step size for each element, we used the adaptive RKM method to integrate phase shifts. We found that in terms of computational time, the RKM solver performs better than a Numerov type scheme, but only in the case where the potential is not too stiff. This usually holds for light elements. In a Yukawa parametrization for example, if $V(r) = Ze^{-ar}/r$, we found that for typically $Z<20$, having a relatively high tolerance, for example $\textrm{tol} = 10^{-3}-10^{-2}$, did not affect greatly the phase shifts. However, as can be seen in Fig. ~\ref{fig:numerov_rkm} \textbf{(b)}, a low tolerance is needed to correctly integrate the $\delta_l$. For heavy elements, for example $\textrm{Au}$, as shown in Fig.~\ref{fig:sigma_dilue} \textbf{(a)}, the rapidly varying potential becomes highly non linear, which makes the error estimate inaccurate, and thus requires low tolerance, to correctly integrate the phase shifts.

To correct for the behavior at large $Z$, we impose a supplementary condition in the RKM scheme. By integrating Eq.~(\ref{cal}) from $r=0$ to $r=+\infty$, we get (with $\delta_\ell(0) = 0$)
\begin{equation}
\delta_\ell(p) = -\int_{0}^{+\infty}\frac{2V(r)}{p}\left[\cos(\delta_\ell)j_\ell(pr)-\sin(\delta_\ell)n_\ell(pr) \right]^2 \mathrm{d}r.
\label{eqcond}
\end{equation}
Now, since $k_1 = h~f(r,\delta_\ell(r))$ and $k_5 = h~f(r + h,\delta_\ell^{\textrm{4'th order}})$, we could write, 
\begin{equation*}\label{eq:rkm_corr}
k_1 + k_5 \approx \int_{r}^{r+h} f(r',\delta_\ell(r'))\mathrm{d}r'.
\end{equation*}
By summing at each step the $k_1$ and $k_5$ coefficients, which does not impact greatly the computation time, we can check if the final phase shift is different from the sum of all the weights. If not, one could locally decrease the tolerance, to better integrate the phase shift in a given region, as in Fig.~\ref{fig:sigma_dilue} \textbf{(b)}, where the result of integration with this adaptative $\textrm{tol}$ is represented.
\subsection{The Born approximation for a Yukawa potential}\label{subsec33}
While integrating the phase shifts, could be achieved either with the RKM solver or a Numerov type scheme, in some cases, one need to sum up to $\ell\approx20$, for the conductivity to converge. In order to explore simplifications, especially at higher values of $\ell$, we seek approximations to Eq.~(\ref{cal}). For example, if the phase shift is assumed small, one immediately gets, with $\delta_{\ell}(0) = 0$, from integrating the Calogero equation,
\begin{equation*}
\delta_{\ell} =-\int_{0}^{\infty}\frac{2V(r)}{p} j_{\ell}^2(pr) \mathrm{d}r.
\end{equation*}
This is the so-called Born approximation. For small phase shifts, one could compute the above integral, however, when the potential is of Yukawa type, or (with $a,b>0$):
\begin{equation}\label{yuk}
V(r) = -a\frac{e^{-br}}{r},
\end{equation}
the phase shift reduces to \cite{Gradshteyn2007},
\begin{equation*}
\delta_{\ell} = \frac{a}{p}Q_{\ell}\left(1+\frac{b^2}{2p^2} \right),
\end{equation*}
where $Q_{\ell}$ is the Legendre function of second kind~\cite{Abramowitz1964}.
\begin{figure}
\centerline{\includegraphics[width=0.8\textwidth]{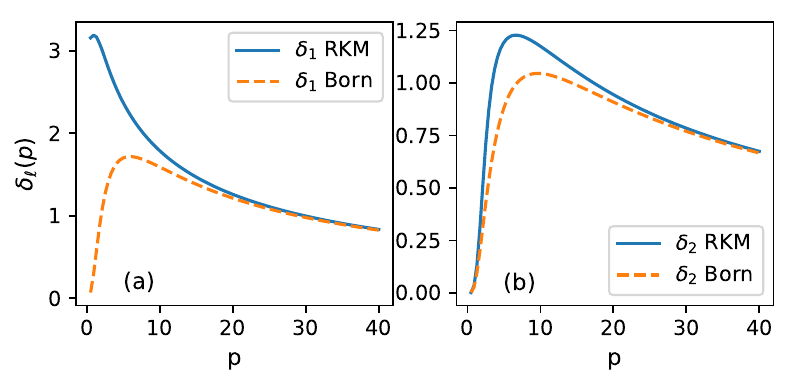}}
\caption{Comparison between $\delta_\ell$ from a numerical resolution of the Calogero equation with the RKM scheme and the Born approximation. In both plots, we use a Yukawa-type potential of $V(r) = -28~e^{-3r}/r$ and an analytical formula for the Born approximation. We plot $\delta_1$ in \textbf{(a)} and $\delta_2$ in \textbf{(b)}.\label{fig:rkm_born}}
\end{figure}
This approximation is especially useful when computing scattering cross sections, since phase shifts decrease with $\ell$, as can be seen on Fig.~\ref{fig:rkm_born}. The Born approximation improves thus with increasing $\ell$. Above a certain $\ell$ it is possible to replace actual scattering phase shifts by Born phase shifts~\cite{Faussurier2019}.

However, one should be careful when conducting this replacement, since different, non equivalent approaches are available to compute the scattering cross-sections and thus conductivities. Let's first explore the classical Born approximation. With a unity structure factor, the Ziman conductivity writes, 
\begin{equation}
\sigma_{\mathrm{Ziman}}(\omega) = \frac{\beta}{3\pi^2} \int_{0}^{\infty}\frac{p^4\nu_p}{\nu_p^2+\omega^2}f_{FD}(1-f_{FD})\mathrm{d}p,
\end{equation}
where $\nu_p =1/\tau_p$ is the relaxation frequency. For the Ziman-Evans conductivity (Z-E), we have $\sigma_{\mathrm{Z-E}}(\omega) = \sigma_0 / (1+\omega^2\sigma_0^2/n_i^2{Z^{*}}^2)$, where  
\begin{equation}\label{sigma0}
\sigma_0 = \frac{3\pi^2{Z^{*}}^2n_i^2}{\beta \int_{0}^{\infty}p^3\nu_pf_{FD}(1-f_{FD})\mathrm{d}p},
\end{equation}
is the DC conductivity. In the two cases, we have to compute the relaxation frequency $\nu_p$, or equivalently, the transport cross-section, $\sigma_{tr}$ in order to obtain the conductivity. For that, one could simply replace in the sum, the phase shifts by Born phase shifts. We have, 
\begin{equation}\label{sig}
\sigma_{tr}^{\mathrm{Born}} = \frac{4\pi}{p^2}\sum_{\ell=0}^{+\infty} (\ell+1)\sin^2(\delta_{\ell+1}^{\mathrm{Born}}-\delta_\ell^{\mathrm{Born}}).
\end{equation}
Alternatively, the transport cross-section could be obtained from the scattering amplitude $f(\theta)$, that is given by,
\begin{equation}
f(\theta) = \frac{1}{p}\sum_{\ell=0}^{\infty}(2\ell+1)e^{i\delta_\ell}\sin(\delta_\ell)P_\ell(\cos(\theta)).
\end{equation}
For that, one simply computes, 
\begin{equation*}
\sigma_{tr} = 2\pi \int_{0}^{\pi} (1-\cos(\theta))\sin(\theta)|f(\theta)|^2\mathrm{d} \theta,
\end{equation*}
where the scattering amplitude could be written in the Born approximation of small phase shifts,
\begin{equation}\label{ftheta}
f_{\mathrm{Born}}(\theta)\approx \frac{1}{p}\sum_{\ell=0}^{\infty}(2\ell+1)\delta_\ell P_\ell(\cos(\theta)).
\end{equation}
We already see that this is a different approximation than previously, since we approximate $\sin(x)\approx x$ and $e^{ix}\approx 1$. However, pursuing with the expression of the phase shifts found earlier, we get,
\begin{equation*}
f_{\mathrm{Born}}(\theta) = \frac{-2}{p^2}\int_{0}^{\infty} V(r) \sum_{\ell=0}^{+\infty} (2\ell+1) j_\ell^2(pr) P_\ell(\cos(\theta)).
\end{equation*}
The summation over $\ell$ simplifies into
\begin{equation*}
\sum_{\ell=0}^{+\infty}(2\ell+1)j_\ell^2(pr)P_\ell(\cos(\theta)) = p^2r^2\frac{sin(qr)}{qr},
\end{equation*}
where $q = 2p\sin(\theta/2)$. The scattering amplitude can then be recast into
\begin{equation*}
f_{\mathrm{\mathrm{Born}}}(\theta) = \frac{-2}{q}\int_{0}^{\infty} rV(r)\sin(qr) dr.
\end{equation*}
and the transport cross-section becomes
\begin{equation}
\sigma^{\mathrm{Born, Fourier}}_{tr}(p) = 2\pi\int_{0}^{\pi}(1-\cos(\theta))\sin(\theta) |f_{\mathrm{Born}}(\theta)|^2 d\theta.
\end{equation}
The change of variable $q \rightarrow 2p\sin(\theta/2)$ enables us to write 
\begin{equation*}
\sigma^{\mathrm{Born, Fourier}}_{tr}(p) = \frac{4\pi}{p^4}\int_{0}^{2p}q \left(\int_{0}^{+\infty}rV(r)\sin(qr) dr\right)^2 dq.
\end{equation*}
In order to avoid numerical integration, to be the most precise possible in our comparison between these two transport cross sections, we find an analytic expression of it, in the case of a Yukawa-type potential $V(r) = -ae^{-br}/r$ (with $a,b>0$). One gets then,
\begin{align}\label{regular_Born}
\sigma^{\mathrm{Born , Fourier}}_{tr}(p) &= \frac{4\pi a^2}{p^4}\int_{0}^{2p}q \left(\int_{0}^{+\infty}e^{-br}\sin(qr) dr\right)^2 dq \\ 
&= \frac{4\pi a^2}{p^4}\int_{0}^{2p}q \left(\frac{q}{b^2+q^2}\right)^2 dq = \frac{4\pi a^2}{p^4}\int_{0}^{2p}\frac{q^3}{(b^2+q^2)^2} dq,
\end{align}
and thus, since
\begin{equation*} 
\frac{x^3}{(b^2+x^2)^2} = \frac{1}{2} \left( \frac{b^2}{b^2+x^2} + \ln(b^2+x^2) \right)',
\end{equation*}
we deduce the analytical expression of the transport cross section.
\begin{equation*}
\sigma_{tr,\mathrm{Born}}^{\mathrm{Fourier}}(p) = \frac{2\pi a^2}{p^4} \left(\ln\left(1+\frac{4p^2}{b^2}\right) - \frac{4p^2}{b^2+4p^2}\right).
\end{equation*}
\begin{figure}
\centerline{\includegraphics[width=0.8\textwidth]{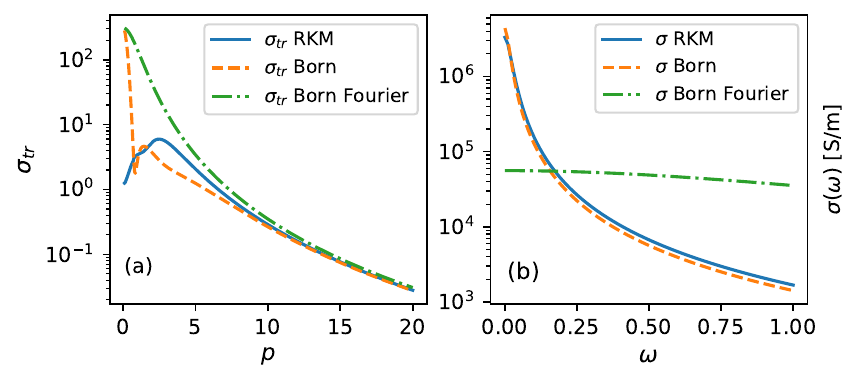}}
\caption{In \textbf{(a)} the numerical scattering cross section $\sigma_{tr}$ is compared with the regular Born approximation given by Eq.~(\ref{regular_Born}), and with the cross section given by Eq. (\ref{sig}) in which analytic Born phase shifts were used in the sum. In \textbf{(b)} we plot the real conductivity in which we used the different cross sections represented in \textbf{(a)}. The two cases are for $V(r) = -28~e^{-3r}/r$.}\label{fig:sigma_tr}
\end{figure}

As can be seen in Fig. \ref{fig:rkm_born}, the Born approximation, expected to be valid at high $p$, provides a fairly good approximation of $\delta_1$ and $\delta_2$ over a wide range. At low $p$, a significant disagreement is observed for $\delta_1$, but even if the Born approximation tends to systematically overestimate the phase shift, the agreement is much better for $\delta_2<\delta_1$. As concerns the transport cross-section (see Fig. \ref{fig:sigma_tr}, \textbf{(a)}), both Born approximations (i.e., the regular Born approximation given by Eq.~(\ref{regular_Born}), and the cross section given by Eq.~\ref{sig} with analytic Born phase shifts) overestimate the values at small $p$, as compared to the RKM numerical results. The analytic Born formulation exhibits a more regular behavior. However, when considering the frequency-dependent conductivity itself, one can notice (see Fig. \ref{fig:sigma_tr}, \textbf{(b)}), that the regular Born approximation is almost superimposed with the RKM values, whereas the analytic Born is very discrepant.

\section{Comparison with \textit{ab initio} calculations}\label{sec4}

\subsection{DFT-MD simulations with ABINIT}\label{subsec41}

To compute the conductivities for $\textrm{Cu}$ and $\textrm{Al}$, we relied on molecular dynamics (MD) simulations. These simulations were performed in the isokinetic ensemble, a framework ensuring that at each step in the Verlet algorithm (with which the ions are advanced each time step) the kinetic energy is conserved. We adapted the time step at the beginning of our simulation, so that the trajectories of each atom were properly sampled, with respect to their mean free path. For $\textrm{Al}$ simulations, we used $32$ atoms initially in an $\textrm{FCC}$ configuration and for \textrm{Cu}, we used 108 atoms in an \textrm{FCC} configuration too. Both $\textrm{Al}$ and $\textrm{Cu}$ are liquid in all of the considered cases, so that the initial lattice type is not important for the molecular dynamics. The $\textrm{Al}$ pseudopotential had $10$ electrons frozen in the core, which is sufficient even for the case at $ \rho_0/10$ and $k_BT=15$ eV, since the estimated $Z^{*}$ is about $2.8$. The $\textrm{Cu}$ pseudopotential contained $19$ frozen electrons in the core.

Unlike ions which are treated classically, the electronic density is self-consistently computed at each time step using the Kohn-Sham density functional theory (KS-DFT) with the Mermin finite temperature functional form. We used specifically designed projector augmented-wave (PAW) pseudo-potentials, already used for the CPTCCW workshop (see Refs. \cite{Stanek2024,Blanchet2024}). We also used short cutoff radii to prevent the PAW spheres from overlapping including most electrons in the valence. To sample the Brillouin zone, we used the Baldereschi mean-value point \cite{Baldereschi1973} for $\textrm{Al}$ and a Monkhorst-Pack grid \cite{Monkhorst1976,Monkhorst1977} for $\textrm{Cu}$, for which we ensured convergence by taking a $2\times2 \times2$ and then a $8\times 8\times 8$ grid.
\begin{figure*}[!ht]
\centerline{\includegraphics[width=\textwidth]{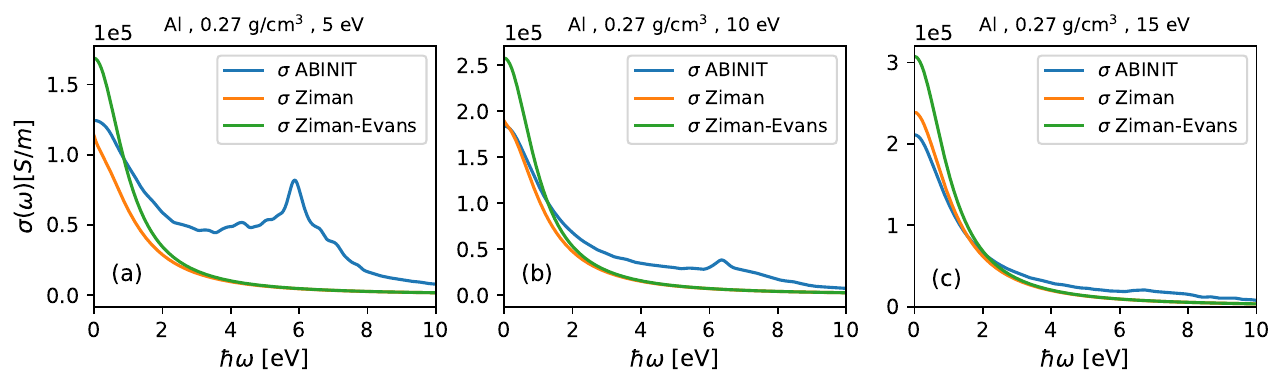}}
\caption{In \textbf{(a)} the conductivity $\sigma(\omega)$ as a function of the frequency $\omega$ in eV from \textit{ab initio} simulations with ABINIT and average atom computation with Ziman and Ziman-Evans models for Al at a tenth of standard density and $k_BT$=5 eV. In \textbf{(b)} the same plot but for $k_BT$=10 eV and in \textbf{(c)} for $k_BT$=15 eV. In all of these plots a hard sphere structure factor was used.\label{fig:sigma_Al}}
\end{figure*}
In order to compute the conductivities and the opacities, from the at least $3000$ time steps long simulations (to ensure proper statistics), we had to select several ionic configurations from the molecular dynamics trajectories and perform more converged DFT calculations. For instance, we increased the Brillouin sampling grid and the number of Kohn-Sham orbitals to include a large number of weakly occupied states. We then applied a Kubo-Greenwood formalism to the eigenstates and eigenvalues to extract the real part of the electrical conductivity. The density of states (DOS) was computed for each of the $6$ snapshots retained of the molecular dynamics simulation and
averaged after aligning the electronic chemical potential to zero. To transform the discrete eigenvalues, we used normalized Gaussian functions with a half-height width of 0.5 eV typically.

\subsection{Average-atom simulations of INFERNO type}\label{subsec42}

Our average-atom model is based on Liberman's INFERNO atom-in-jellium model of matter \cite{Liberman1979}. The T and A versions of the model differ in the method used to calculate thermodynamic quantities. In the T model, quantities related to the jellium are systematically subtracted from the quantities calculated across the entire space. In the A model (considered here), the separation between the ionic cell and the jellium is spatial in nature: thermodynamic quantities are calculated only within the ionic cell. A key assumption of the model is the muffin-tin approximation, which posits that beyond the ionic cell, the electron density remains constant and equals the jellium density. This assumption, derived from solid-state theory, has significant implications for the model. It simplifies the system's global neutrality to the neutrality of the ionic cell, ensuring that the potential is zero outside the cell. Consequently, the atomic sphere is effectively isolated from the surrounding jellium, and all calculations are restricted to the ionic (Wigner-Seitz) cell. Electronic structure is then computed by a self-consistent iterative scheme, providing the energies and average electron populations of the subshells, together with the mean-field electrostatic potential and the chemical potential. The only needed parameters are the atomic number $Z$, density $\rho$, and temperature $T$. Average-atom methods are well-regarded for providing fairly accurate results at a reasonable computational cost.

\subsection{Ziman and Ziman-Evans comparison with a structure factor}\label{subsec43}

In order to compare the Ziman and Ziman-Evans formulas, we chose first the case of $\textrm{Al}$ at $\rho_0/10$ and different temperatures, presented on Fig.~\ref{fig:sigma_Al}. Since the system is dilute and relatively warm, the ion-ion correlation function should be close to a Heaviside function $g(r) = 0$ for $r\le r_{\mathrm{ws}}$, and $g(r)=1$, for $r>r_{\mathrm{ws}}$, which is confirmed by our \textit{ab initio} simulations. This ion-ion correlation function is consistent with the ion-sphere (confined atom) model. It is thus relevant to resort to the so-called hard-sphere structure factor. The structure factor being the radial Fourier transform of $g(r)$, we obtain:
\begin{equation*}
    S(q)=1+4\pi n_i\int_0^{\infty}r^2g(r)\frac{\sin(qr)}{qr}\,\mathrm{d}r,
\end{equation*}
and we get
\begin{equation*}
    S(q)=1-\frac{3}{(qr_{\mathrm{ws}})^3}\left[\sin(qr_{\mathrm{ws}})-qr_{\mathrm{ws}}\,\sin(qr_{\mathrm{ws}})\right], 
\end{equation*}
where $r_{\mathrm{ws}}$ is the radius of the Wigner-Seitz sphere related to the ionic density $n_i$ by
\begin{equation*}
    \frac{4}{3}\pi r_{\mathrm{ws}}^3\,n_i=1.
\end{equation*}
\begin{figure}[!ht]
\centerline{\includegraphics[width=0.8\textwidth]{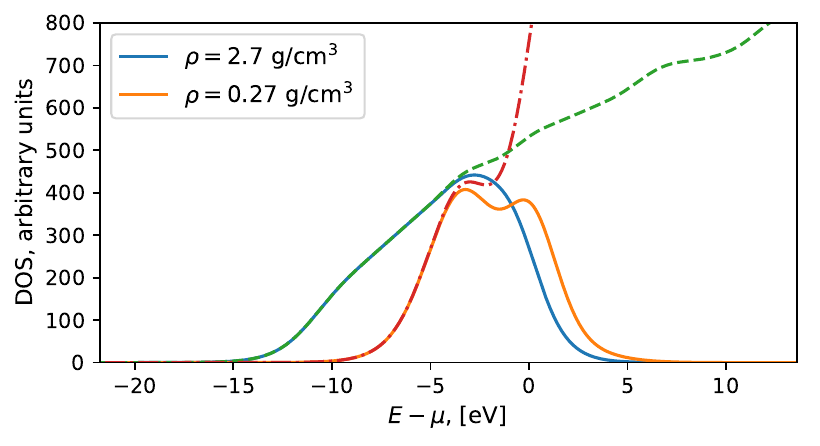}}
\caption{The electronic density of states (DOS) of Al at solid density and a tenth of solid density as a function of the electron's energy, the chemical potential substracted. The temperature is such that $k_BT$=1 eV. The full solid curves represent the occupied DOS, multiplied by the Fermi-Dirac distribution. The dashed and dashed-dotted lines represent the free DOS. We see the $3s$ and $3p$ states at $\rho_0/10$ which are merged into a broader structure at higher density.\label{fig:Al_DOS}}
\end{figure}
As can be seen in Fig.~\ref{fig:sigma_Al}, there is a quite good agreement between the Ziman frequency-dependent conductivity and the DFT-MD simulations. We also observe that it becomes increasingly better with the temperature rise. For $k_BT = 5$ eV, we see a pronounced peak in the conductivity near $6$ eV, which correspond to semi-bound transitions from the $3s$ and $3p$ states, which are merged into a broad structure at $\rho_0$, as can be seen on the density of states (DOS) on Fig.~\ref{fig:Al_DOS}. For $k_BT = 10$ eV, the peak is reduced, because of the broadening of the DOS with temperature. At $k_BT=15$ eV, the peak is even more reduced, and the agreement of the now almost only free-free conductivity is better than in the lower temperature cases.
As for the differences between the Ziman and Ziman-Evans formulas, a part from different formulations, it can come from the mean ionization $Z^{*}$ additionally present in the Ziman-Evans formulation. The difference is mainly present for the DC conductivity, at large frequencies, both models reducing mainly to a Drude form.

\begin{figure*}[ht!]
\centerline{\includegraphics[width=\textwidth]{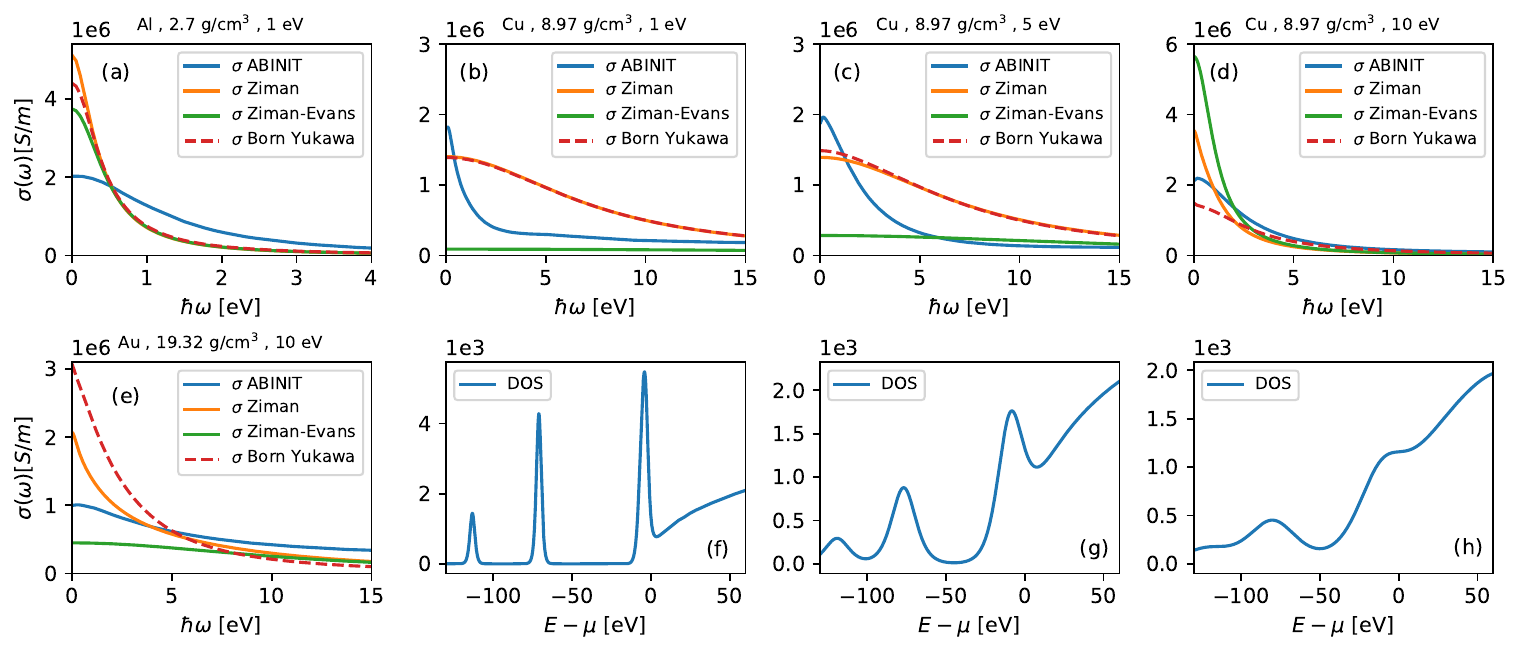}}
\caption{In plots \textbf{(a)} to \textbf{(e)} we represent the results of our \textit{ab initio} conductivity computations along with the Ziman and Ziman-Evans models, with unity structure factor. We also give a fully analytical formula, represented by the dashed line, which is based on Born phase shifts and a Yukawa-type fit of the electron-ion potential. The plots \textbf{(f)} to \textbf{(g)} are the density of states (in arbitrary units) corresponding to the plots \textbf{(b)} to \textbf{(d)}. These DOS were computed using DFT based MD simulations using ABINIT and then averaged on snapshots after substracting the chemical potential $\mu$ of each snapshot.\label{fig:result}}
\end{figure*}

Now, for the WDM regime, we mainly chose cases already presented at the CPTCCW workshop, as that of $\textrm{Al}$, $\textrm{Cu}$ and $\textrm{Au}$ at $\rho_0/10$ and $k_BT=1$ eV for the first two and $10$ eV for the last one. For $\textrm{Cu}$, we also computed the cases $k_BT=5$ eV and $k_BT=10$ eV in order to be able to discuss the effect of the DOS broadening on the conductivity. The results are presented on Fig.~\ref{fig:result}. A unity structure factor was used for the simulations presented in this plot, since a Heaviside ion-ion correlation function does not describe all of these conditions. Comparisons with more evolved structure factors was left for future work.

For $\textrm{Al}$ at $k_BT=1$ eV, consistent with other average-atom computations presented at the CPTCCW workshop, the Ziman DC conductivity is more than twice the \textit{ab initio} conductivity. However, since there are no localized electrons on the DOS corresponding to this case, as can be seen in Fig.~\ref{fig:Al_DOS}, the conductivity could still be described with a Drude formula. This is not the case for $\textrm{Cu}$ at $k_BT=1$ eV, as was shown in~\cite{Blanchet2024}, and is visible on Fig.~\ref{fig:result} \textbf{(b)}. The reason is due to electrons localized on the $d$-shell, as can be seen on the DOS on Fig.~\ref{fig:result} \textbf{(f)}. Here, again, the poor agreement between the Ziman-Evans formula and the Ziman formula, may come from an inaccurate prediction of the mean ionization $Z^{*}$ by average atom models. As the temperature increases, for example $k_BT=5$ eV and $k_BT=10$ eV, the temperature broadening of the DOS, visible on Fig.~\ref{fig:result} \textbf{(g)-(h)}, translates directly into the smoothing of the frequency-dependent conductivity on plots Fig.~\ref{fig:result} \textbf{(c)-(d)}. In particular, conductivities become more Drude-like. Also, with the increase of $Z^{*}$, the Ziman-Evans model's predictions, becomes more accurate, particularly at high values of $\hbar\omega$. For the $\textrm{Au}$ case, we see that both high energy predictions and the DC conductivities could at best give a qualitative agreement. 

We would also like to draw the reader's attention on the red dashed lines, that represent the Born-Yukawa analytical formula, where the Born phase shifts were directly included in the transport cross section, and where a Yukawa fit of the average-atom electron-ion potential was used. The formula is quite close in these situations to the Ziman conductivities, and can be used for example as a first attempt, to compute low frequency opacity tables.
\begin{figure}[ht!]
\centerline{\includegraphics[width=0.8\textwidth]{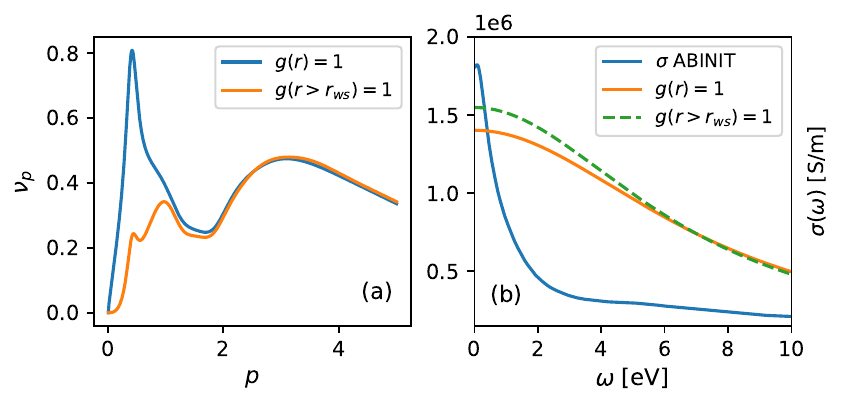}}
\caption{The two subplots correspond to copper at solid density and $k_BT$=1 eV. In \textbf{(a)}, we see the relaxation frequency as a function $\nu_p = 1/\tau_p$ as a function of the momentum $p$ in a case without structure factor (so $g(r)= 1$) and with a hard spheres structure factor, when $g$ is a Heaviside function of span $r_{\mathrm{ws}}$. On the \textbf{(b)} image, we see the conductivity in the same configuration, with and without structure factor, compared with results from \textit{ab initio} simulations. \label{fig:structure}}
\end{figure}

\begin{figure}[ht!]
\centerline{\includegraphics[width=0.8\textwidth]{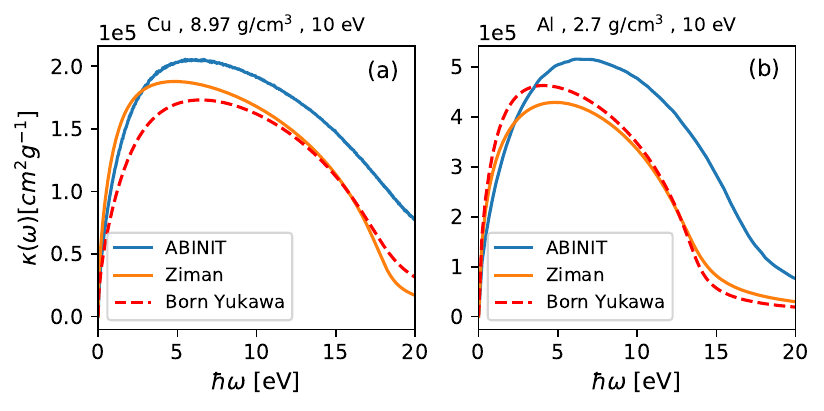}}
\caption{The free-free opacity of $\textrm{Cu}$ at solid density and $k_BT$= 10 eV as a function of the transition energy in eV. A comparison is made with the opacity obtained from Ziman conductivity and with an analytical formula, which is the Born-Yukawa approximation. In \textbf{(b)} - the same plot, but for $\textrm{Al}$ at solid density and $k_BT$=10 eV. \label{fig:opacity}}
\end{figure}

As for the structure factor, more evolved ones, such as an one-component-plasma (OCP)~\cite{Rinker1985, Bretonnet1988}, could improve the agreement with DFT-MD simulations. In our case, we only tested the hard spheres structure factor, as for example on Fig.~\ref{fig:structure}. Here we see that, the inclusion of a structure factor, which does not precisely describe the correlated case of $\textrm{Cu}$ at solid density and $k_BT=1$ eV, already improve the agreement betweeen the Ziman DC conductivity and the $\textit{ab initio}$ computation.
As explained in \ref{subsec24}, from the frequency-dependent conductivity, one can compute the frequency-dependent opacity via the optical index. On Fig.~\ref{fig:opacity}, we report frequency-dependent opacities for $\textrm{Cu}$ and $\textrm{Al}$ at $\rho_0$, and $k_BT=10$ eV. In both cases, at high temperatures, the DOS broadening ensures a Drude-like form of the conductivity, making it easier for the Ziman formalism to reach an agreement with \textit{ab initio} simulations. In this case, opacities were computed directly with ABINIT, which uses the same procedure as that outlined in \ref{subsec24}.

\subsection{Verifying the sum rule}\label{subsec44}

As can be seen in Eq.~(\ref{eta}), the Ziman-Evans formula for the resistivity depends explicitly on the inverse of the square of the mean ionization $Z^*$. Thus, a small change in $Z^*$ has a strong impact on the resistivity. This is particularly troublesome at low temperatures, when $Z^*$ is small, especially since average atom models are questionable and reach their limits under these conditions. This is not the case with the Ziman formula, which does not explicitly depend on $Z^*$, and is {\it de facto} less sensitive to the electronic structure. It does, however, depend on it {\it via} the chemical potential, but there is less uncertainty about this quantity, which makes the Ziman formula more robust in this sense. In general, the sum rule for the conductivity reads \cite{Johnson2006}
\begin{equation}
\frac{2}{\pi}\int_0^{\infty}\sigma(\omega)~\mathrm{d}\omega=n_e=Z^*n_i.
\end{equation}
In the Ziman-Evans formalism, this rule is automatically enforced with the choice of the Drude relaxation time. If $\sigma_{\mathrm{Z-E}}(\omega) = \sigma_0/(1+\omega^2\tau^2)$, then
\begin{equation*}
\frac{2}{\pi}\int_0^{\infty}\sigma_{\mathrm{Z-E}}(\omega)~\mathrm{d}\omega=\frac{\sigma_0}{\tau} = Z^*n_i,
\end{equation*}
and with $\tau = \sigma_0/(Z^{*}n_i)$, the sum-rule is verified automatically. For Ziman's conductivity, verifying the sum-rule would amount to changing the chemical potential, or introducing a supplementary factor or variable. We chose in our case not to enforce it, also because of the difficulty to obtain reliable ionisations in the warm dense matter regime in general, especially with average-atom computations.
\section{Conclusions}\label{sec5}
In this work, we compared two different formalisms to compute the low frequency opacity in the average-atom approximation, the Ziman and the Ziman-Evans models. For this comparison to be more consistent, we explicitly included an ionic structure factor in the Ziman's formula, with the same approximation as the one used to derive the Ziman-Evans conductivity. This approximation consists, in the presence of an array of scatterers, to set the self-consistent solution of the Lippmann-Schwinger equation as being that of a single scatterer inside a given non overlapping potential well. The presence of a shift in the incoming free particle wave, due to the displacement of each given scatterer from the origin, allows naturally to account for the ionic structure factor. 

For the two models, we had to evaluate scattering phase shifts, that we computed from Calogero equation with an adaptive step numerical scheme. The adaptive scheme was that of Runge-Kutta-Merson, based on a classical RK4 scheme, with a supplementary evaluation, allowing for an adaptive step. We compared this scheme with the usual Numerov one for the Schr\"odinger equation and found that it performed better, but only for sufficiently light elements, such as $Z<20$. For heavier elements, we gave some ideas on how to improve the Runge-Kutta-Merson scheme. 

We also explored different approximations, such as Born approximation for phase shifts, and showed that approximating the phase shifts instead of the total scattering cross section gave more accurate results, by a order of magnitude in some cases. 

The average-atom results were compared with \textit{ab initio} simulations first for $\textrm{Al}$ at $\rho_0/10$ and different temperatures and then for $\textrm{Cu}$, $\textrm{Al}$ and $\textrm{Au}$ in the warm dense matter regime. As expected, the authors found agreement between the DFT-based MD simulations and the average atom computations, for dilute aluminum. The authors also found that increasing the temperature for transition metals, led to a more accurate description of the transport properties such as of the conductivity and opacity. Future work is necessary though, to expand the average-atom models to temperatures such that $k_BT<10$ eV, especially for transition metals.

\section{Appendix A: Resistivity in the low ion density limit from Kubo formula}\label{appA}

First, we derive the Kubo formula, in the usual current-current correlation form, and then in momentum-position form. This allows one to directly take the $\omega \to 0$ limit in the integral. We consider a system of $N_e$ non-interacting electrons, and $N_s$ scatterers, with the Hamiltonian given by $\hat{H}_0$ in Eq.~(\ref{eq:hamiltonian}). 
At thermal equilibrium, in the grand-canonical ensemble of unit volume, temperature $T$, and chemical potential $\mu$, the density matrix operator can be written as $\hat{\rho}_0 = e^{-\beta(\hat{H}_0-\mu\hat{N})}/Z_0$ with $\beta = 1/ k_BT$ and $Z_0 = \textrm{Tr}[e^{-\beta(\hat{H}_0-\mu \hat{N})}] = \sum_{n}e^{-\beta (E_n-\mu N_n)}$. Here $\hat{N}$ is the number operator and $N_n$ is the occupation number. We also have $\hat{H}_0|n\rangle = E_n|n\rangle$, $E_n$ being the eigenvalues of $\hat{H}_0$ and $|n\rangle$ being its eigenvectors.

In the presence of an electric field, say $\textbf{E} = E_0e^{-i\omega t}\textbf{y}$, the Hamiltonian $\hat{H}_0$ becomes, 
\begin{equation*}
\hat{H} = \frac{1}{2}\left(\hat{\textbf{p}}-\hat{\textbf{A}}\right)^2 + \hat{V}(\textbf{r}) = \hat{H}_0 + \hat{H}_1,
\end{equation*}
with $\hat{H}_1 = (\hat{\textbf{p}} \cdot \hat{\textbf{A}} + \hat{\textbf{A}}\cdot\hat{\textbf{p}})/2 + \hat{A}^2/2$. If we introduce the current operator $\hat{\textbf{j}}$, then the perturbation $\hat{H}_1$ writes as, $\hat{H}_1 = \hat{\textbf{j}}\cdot\hat{\textbf{A}} + \hat{A}^2/2 $. If we write $\textbf{A} = -i/\omega \textbf{E}$, we readily find $\hat{H}_1 = -i/\omega \hat{\textbf{j}}\cdot\hat{\textbf{E}} - \hat{E}^2/2\omega^2 $ The next step is to write the Liouville equation, describing the time evolution of the matrix operator $\hat{\rho}(t) = \hat{\rho}_0 + \hat{\rho}_1$ as,
\begin{equation*}
\frac{\partial \hat{\rho}}{\partial t} = -i[\hat{H},\hat{\rho}].
\end{equation*}
From this, we find at the first order that, $\partial_t \hat{\rho}_1 = -i[\hat{H}_1,\hat{\rho}_0]-i[\hat{H}_0,\hat{\rho}_1]$. It can be checked then, by direct differentiation, that the solution to the linearized equation is given by, 
\begin{equation*}
\hat{\rho}_1(t) = -i\int_{-\infty}^{t}e^{-i(t-t')\hat{H}_0}\left[ \hat{H}_1(t'),\hat{\rho}_0\right]e^{i(t-t')\hat{H}_0}\mathrm{d}t',
\end{equation*}
with $\hat{\rho}_1(-\infty) = 0$. It is implicitly assumed here that the perturbation is turned on adiabatically, which means that the electric field grows from $t=-\infty$, as $e^{\eta t}$, with $\eta \to 0$.

The average value, of an operator $\hat{O}$ could then be computed with $\langle\hat{O}\rangle = \textrm{Tr}(\hat{\rho}\hat{O}) = \textrm{Tr}(\hat{\rho}_0\hat{O}) + \textrm{Tr}(\hat{\rho}_1\hat{O}) = \langle\hat{O}\rangle_0 +\langle\hat{O}\rangle_1 $. Usually, the average value computed with the unperturbed matrix vanishes in the absence of external fields. It is the case with the current operator $\hat{\textbf{j}}$ and also with the force operator $\hat{\textbf{F}}$, which we recall as being $\hat{\textbf{F}} = i[\hat{H},\hat{\textbf{p}}]$. If we compute for example the average value of the current operator $\langle\hat{\textbf{j}}\rangle = \textrm{Tr}(\hat{\rho}_1\hat{\textbf{j}})$, using the expression we found for $\hat{H}_1$, we obtain the following formula,
\begin{equation*}
\langle \hat{\textbf{j}}(t) \rangle = \frac{1}{\omega}\int_{-\infty}^{t}\textrm{Tr}\left(e^{-i(t-t')\hat{H}_0}\left[ \hat{\textbf{j}}(t'),\hat{\rho}_0\right]e^{i(t-t')\hat{H}_0}\hat{\textbf{j}}(t)\right)\textbf{E}(t')\mathrm{d}t' + i\textbf{E}/\omega.
\end{equation*}
Here, $i\textbf{E}/\omega$ is the diamagnetic term. The conductivity $\sigma(\omega)$ is found after a Fourier transform of the previous relation and with the cyclic property of the trace operator to be,
\begin{equation*}
\sigma(\omega) = \frac{1}{\omega}\int_{0}^{+\infty}e^{i\omega t}\langle[ \hat{\textbf{j}}(t),\hat{\textbf{j}}(0)]\rangle\mathrm{d}t + \frac{i}{\omega}.
\end{equation*}
An integration by parts of the conductivity, with operators projected on the $\textbf{x}$ axis, (for example $\hat{j}_x = \hat{p}_x$ in atomic units) and with $d p_x/dt = i[\hat{H}_0,\hat{p}_x]$, gives Eq.~(\ref{eq:kubo1}), after the cyclic property of the trace is used again,
\begin{equation*}
\sigma(\omega) = i\int_{0}^{+\infty} \mathrm{d}t e^{i\omega t} \langle [\hat{p}_x(t),\hat{x}(0)]\rangle.
\end{equation*}
From this formula we see that if $\omega = 0$, $\sigma(0) = i\int_{0}^{+\infty}\langle[\hat{p}_x(t),\hat{x}(0)]\rangle \mathrm{d} t$. However, as said previously, this DC conductivity diverges if the density of scatterers goes to $0$.

To derive the resistivity formula, we perform two integrations by parts and use regular properties of quantum mechanical operators, to obtain,
\begin{equation*}
\sigma(\omega) = \frac{i}{\omega} + \frac{1}{\omega^2} \int_{0}^{+\infty} \mathrm{d}t e^{i\omega t} \int_{0}^{\beta}\mathrm{d}\lambda \langle \hat{F}_x(0)\hat{F}_x(t+i\lambda)\rangle.
\end{equation*}
Here we have used the following identity, which holds for any $\hat{B}$,
\begin{equation*}
[e^{-\beta \hat{H}_0},\hat{B}] = e^{-\beta \hat{H}_0}\int_{0}^{\beta}\mathrm{d}\lambda e^{\lambda \hat{H}_0} [\hat{H},\hat{A}]e^{-\lambda \hat{H}_0}.
\end{equation*}
The resistivity is then given by $R(\omega) = 1/\sigma(\omega)$, which can be expanded at first order in the scatterer density $n_s$ (or in the scattering amplitude) as follows, 
\begin{equation*}
R(\omega) = \int_{0}^{+\infty} \mathrm{d}t e^{i\omega t} \int_{0}^{\beta}\mathrm{d}\lambda \langle \hat{F}_x(0)\hat{F}_x(t+i\lambda)\rangle + O(n_s^2).
\end{equation*}
This formula could be evaluated on a complete set of diffusion states (solutions to Lippmann-Schwinger equation) such that $\int \mathrm{d}^3k |k\rangle\langle k| = 1$ and $\hat{H}_0|k\rangle = k^2|k\rangle$. This base is complete provided that there are no bound states\cite{Roman1965}. However, we find it more instructive (and simple in terms of notations) to treat the discrete case, the continuous case being a direct generalization. We consider, as previously a system composed of $N_e$ non interacting electrons, described by their occupation numbers $n_i$, and labeled in the Fock space as $|i_1,...,i_{N_e}\rangle$. This is a basis which diagonalizes the operator $\hat{H}_0-\mu\hat{N}$, such that, 
\begin{equation*}
\langle i_1,...,i_{N_e}|(\hat{H}_0-\mu\hat{N})|i_1,...,i_{N_e}\rangle = \sum_{i=1}^{N_e}(\varepsilon_i-\mu)n_i.
\end{equation*}
To simplify notations, we denote by $|i_{1...N_e}\rangle$ the ket describing the $N_e$ non interacting electrons. For the average value in the integral, we write the trace as sums over occupation numbers (which for fermions can only take values in $\{0,1\}$),
\begin{equation*}
\begin{aligned}
Tr[\hat{\rho}_0\hat{F}_x(0)\hat{F}_x(t+i\lambda)] &= \sum_{i_1} ... \sum_{i_{N_e}} \langle i_{1...N_e}|\hat{\rho}_0\hat{F}_x(0)\hat{F}_x(t+i\lambda)|i_{1...N_e}\rangle.
\end{aligned}
\end{equation*}
A closure relation could be introduced between the force operators as follows, 
\begin{equation*}
\begin{aligned}
Tr[\hat{\rho}_0\hat{F}_x(0)\hat{F}_x(t+i\lambda)] &= \sum_{i_1} ... \sum_{i_{N_e}} \sum_{j_1} ... \sum_{j_{N_e}} \langle i_{1...N_e}|\hat{\rho}_0\hat{F}_x(0)|j_{1...N_e}\rangle \\
&\times\langle j_{1...N_e}|\hat{F}_x(t+i\lambda)|i_{1...N_e}\rangle.
\end{aligned}
\end{equation*}
With the help of the definition of the density matrix, we can evaluate the bra-ket's as, 
\begin{equation*}
\langle i_{1...N_e}|\hat{\rho}_0\hat{F}_x(0)|j_{1...N_e}\rangle = \frac{e^{-\beta\sum_{i=1}^{N_e}(\varepsilon_i-\mu)n_i}}{Z}\langle i_{1...N_e}|\hat{F}_x(0)|j_{1...N_e}\rangle,
\end{equation*}
and given the action of the evolution operator, $\hat{F}_x(t+i\lambda) = e^{i(t+i\lambda)\hat{H}_0}\hat{F}_x(0)e^{-i(t+i\lambda)\hat{H}_0}$, so that,
\begin{equation*}
\langle j_{1...N_e}|\hat{F}_x(t+i\lambda)|i_{1...N_e}\rangle = e^{-\lambda(E_j-E_i)} e^{-it(E_i-E_j)}  \langle j_{1...N_e}|\hat{F}_x(0)|i_{1...N_e}\rangle,
\end{equation*}
where $E_i = \sum_{i=1}^{N_e} \varepsilon_i n_i$ is the total energy of the non-interacting electron system described by $|i_1,...,i_{N_e}\rangle$. Let us also denote $N_i = \sum_{i=1}^{N_e}n_i$ - the total occupation number. Then, the DC resistivity can be written as (with $\omega$ having a small positive imaginary part $\eta$), 
\begin{equation*}
\begin{aligned}
R = \int_{0}^{\infty} \mathrm{d}t e^{-\eta t} \int_{0}^{\beta}&\frac{\mathrm{d}\lambda}{Z}\sum_{i_1} ... \sum_{i_{N_e}}\sum_{j_1} ... \sum_{j_{N_e}}e^{-\beta(E_i-\mu N_i)}\, e^{-\lambda(E_j-E_i)} e^{-it(E_i-E_j)}  |\langle i_{1...N_e}|\hat{F}_x(0)|j_{1...N_e}\rangle|^2.
\end{aligned}
\end{equation*}
The time integral, can be written using the real part $\Re$ as $\Re[\int_{0}^{\infty}e^{-it(E_i-E_j-i\eta)}\mathrm{d}t] = \Re[1/i(E_i-E_j-i\eta)] = \pi\delta(E_i-E_j)$, in the limit where $\eta \to 0$. It follows that, 
\begin{equation*}
\begin{aligned}
R = \pi \int_{0}^{\beta}&\frac{\mathrm{d}\lambda}{Z}\sum_{i_1} ... \sum_{i_{N_e}}\sum_{j_1} ... \sum_{j_{N_e}}e^{-\beta(E_i-\mu N_i)}\,e^{-\lambda(E_j-E_i)} \delta(E_i-E_j)  |\langle i_{1...N_e}|\hat{F}_x(0)|j_{1...N_e}\rangle|^2.
\end{aligned}
\end{equation*}
After performing the integral in $\lambda$, we get, 
\begin{equation*}
\begin{aligned}
R = \pi \sum_{i_1} ... \sum_{i_{N_e}}\sum_{j_1} ... \sum_{j_{N_e}}&\frac{e^{-\beta(E_i-\mu N_i)} - e^{-\beta(E_j-\mu N_i)}}{(E_j-E_i)Z} \delta(E_i-E_j)\, |\langle i_{1...N_e}|\hat{F}_x(0)|j_{1...N_e}\rangle|^2.
\end{aligned}
\end{equation*}
Now to continue evaluating this expression, one need to separate it into two sums $R_+$ and $R_-$, such that $R=R_+-R_-$. For example, the first can be written after summing over the two occupation states corresponding to $n_{i_1} = 0$ and $n_{i_1}=1$ as,
\begin{equation*}
\begin{aligned}
R_+ = \pi \sum_{n_{i_1}\in\{0,1\}}\frac{e^{-\beta(\varepsilon_{i_1}-\mu) n_{i_1}}}{Z} \sum_{i_2}... \sum_{i_{N_e}}{e^{-\beta(E_i-\mu N_i)}} &\sum_{j_1} ... \sum_{j_{N_e}} \frac{\delta(E_i-E_j)}{E_j-E_i} \,|\langle i_{1...N_e}|\hat{F}_x(0)|j_{1...N_e}\rangle|^2.
\end{aligned}
\end{equation*}
Since the sum over $n_{i_1}$ readily simplifies as $1+e^{-\beta(\varepsilon_{i_1}-\mu)}$, and since the partition function writes, 
\begin{equation*}
\mathscr{Z} = \sum_{n_1=0}^{1}...\sum_{n_{N_e}=0}^{1} e^{-\beta(\varepsilon_1-\mu)n_1}...e^{-\beta(\varepsilon_{N_e}-\mu)n_{N_e}} = \prod_{i=1}^{N_e}[1+e^{-\beta(\varepsilon_i-\mu)}],
\end{equation*}
we obtain, 
\begin{equation*}
\begin{aligned}
R_+ =&  \frac{\pi}{1+e^{\beta(\varepsilon_{i_1}-\mu)}} 
\sum_{j_1} ... \sum_{j_{N_e}} \frac{\delta(E_i-E_j)}{E_j-E_i} 
|\langle i_{1...N_e}|\hat{F}_x(0)|j_{1...N_e}\rangle|^2 \\
& + \frac{\pi}{\mathscr{Z}} \sum_{i_2}... \sum_{i_{N_e}} e^{-\beta(\varepsilon_{i_2}-\mu) n_{i_2}}...e^{-\beta(\varepsilon_{i_{N_e}}-\mu) n_{i_{N_e}}} \,\sum_{j_1} ... \sum_{j_{N_e}} \frac{\delta(E_i-E_j)}{E_j-E_i} 
|\langle i_{1...N_e}|\hat{F}_x(0)|j_{1...N_e}\rangle|^2,
\end{aligned}
\end{equation*}
where $E_i = \sum_{s=2}^{N_e}\varepsilon_{i_s}n_{i_s}$. For $R_-$, we perform the same transformation, we first isolate the sum over $n_{j_1} \in \{0,1\}$, to obtain,
\begin{equation*}
\begin{aligned}
R_- =&  -\frac{\pi}{1+e^{\beta(\varepsilon_{j_1}-\mu)}} 
\sum_{i_1} ... \sum_{i_{N_e}} \frac{\delta(E_i-E_j)}{E_j-E_i} 
|\langle i_{1...N_e}|\hat{F}_x(0)|j_{1...N_e}\rangle|^2 \\
& + \frac{\pi}{Z} \sum_{i_2}... \sum_{i_{N_e}}  \sum_{j_1} ... \sum_{j_{N_e}}  e^{-\beta[\varepsilon_{j_2}n_{j_2}+...+\varepsilon_{j_{N_e}}n_{N_e}-\mu(n_{i_1}+...+n_{i_{N_e}})]}\,\frac{\delta(E_i-E_j)}{E_j-E_i} 
|\langle i_{1...N_e}|\hat{F}_x(0)|j_{1...N_e}\rangle|^2.
\end{aligned}
\end{equation*}
Here, again $E_j = \sum_{s=2}^{N_e}\varepsilon_{j_s}n_{j_s}$. Let us now show that when summing $R_+$ and $R_-$, the term containing $\pi/Z$ will vanish. First it reads, 
\begin{equation*}
\begin{aligned}
&\frac{\pi}{\mathscr{Z}} \sum_{i_2}... \sum_{i_{N_e}}  e^{-\mu(n_{i_1}+...+n_{i_{N_e}})}\\
&\times\sum_{j_1} ... \sum_{j_{N_e}} \left( e^{-\beta[\varepsilon_{j_2}n_{j_2}+...+\varepsilon_{j_{N_e}}n_{i_{N_e}}]}-e^{-\beta[\varepsilon_{j_2}n_{j_2}+...+\varepsilon_{j_{N_e}}n_{j_{N_e}}]} \right ) \frac{\delta(E_i-E_j)}{E_j-E_i}\,|\langle i_{1...N_e}|\hat{F}_x(0)|j_{1...N_e}\rangle|^2.
\end{aligned}
\end{equation*}
We see that due to the Dirac distribution, the difference of the exponentials is zero. The sum vanishes and we have, 
\begin{equation*}
\begin{aligned}
R =& \frac{\pi}{1+e^{\beta(\varepsilon_{i_1}-\mu)}} 
\sum_{j_2} ... \sum_{j_{N_e}} \frac{\delta(E_i-E_j)}{E_j-E_i} 
|\langle i_{1...N_e}|\hat{F}_x(0)|j_{1...N_e}\rangle|^2 \\
&- \frac{\pi}{1+e^{\beta(\varepsilon_{j_1}-\mu)}} 
\sum_{i_2} ... \sum_{i_{N_e}} \frac{\delta(E_i-E_j)}{E_j-E_i} 
|\langle i_{1...N_e}|\hat{F}_x(0)|j_{1...N_e}\rangle|^2.
\end{aligned}
\end{equation*}
It is understood that when computing the resistivity, more precisely, the trace, we take an integral over the electron energies. Thus, after integrating over the energies of the electrons indexed by $i_2,...,i_{N_e}$ and $j_2, ... , j_{N_e}$, we obtain, after denoting by $\varepsilon_1 = \varepsilon_{i_1}$ and $\varepsilon_2 = \varepsilon_{j_2}$,
\begin{equation*}
\begin{aligned}
R = \pi\int_{0}^{\infty} \mathrm{d}\varepsilon_1 \int_{0}^{\infty}\mathrm{d}\varepsilon_2&\left[\frac{1}{1+e^{\beta(\varepsilon_1-\mu)}}  - \frac{1}{1+e^{\beta(\varepsilon_2-\mu)}} 
 \right]\frac{\delta(\varepsilon_1-\varepsilon_2)}{\varepsilon_2-\varepsilon_1} \,|\langle \varepsilon_1|\hat{F}_x(0)|\varepsilon_2\rangle|^2.
\end{aligned}
\end{equation*}
We dropped the summation on the $i$ and $j$ states, since the bra-kets only depend on the single electron energies that we isolated in the $i$ and $j$ kets. We also see the Fermi-Dirac function $f_{FD}$ derivative appear in the integral. We obtain, 
\begin{equation*}
\begin{aligned}
R = -\pi\int_{0}^{\infty} \mathrm{d}\varepsilon_1 \int_{0}^{\infty}\mathrm{d}\varepsilon_2 f_{FD}'(\varepsilon_1) \delta(\varepsilon_1-\varepsilon_2)|\langle \varepsilon_1|\hat{F}_x(0)|\varepsilon_2\rangle|^2.
\end{aligned}
\end{equation*}
We introduce an integral over an energy variable $\varepsilon$, to obtain, 
\begin{equation*}
\begin{aligned}
R = -\pi\int_{0}^{\infty}\mathrm{d}\varepsilon f_{FD}'(\varepsilon) \int_{0}^{\infty} \mathrm{d}\varepsilon_1 \int_{0}^{\infty}\mathrm{d}\varepsilon_2 \delta(\varepsilon-\varepsilon_1)\delta(\varepsilon-\varepsilon_2)|\langle \varepsilon_1|\hat{F}_x(0)|\varepsilon_2\rangle|^2.
\end{aligned}
\end{equation*}
After averaging over directions and accounting for spin degeneracy, we get the $2/3$ factor and after assuming $N_e$ electrons and $N_s$ scattering centers, with $Z^{*} = N_e/N_s$, we obtain the formula announced in \ref{subsec22}. We recall that at $t=0$, the internal force can be written as $\hat{F}_x(0) = \nabla V(\textbf{r})$, where $V$ is the total electron-ion potential.
\section{Appendix B: Lippmann-Schwinger equation for a single scatterer}\label{appB}
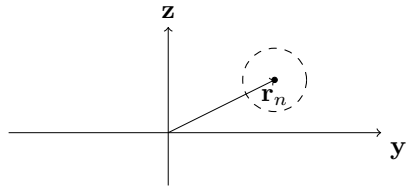
\begin{figure}[!ht]
    \centering
    \begin{tikzpicture}[scale=0.7]
    
    \draw[line width=0.1mm,->] (-3,0) -- (4,0) node[below right] {$\textbf{y}$};
    \draw[line width=0.1mm,->] (0,-1) -- (0,2) node[above] {$\textbf{z}$};
    \draw[line width=0.1mm,->] (0,0) -- (2,1) node[below] {$\textbf{$\textbf{r}_n$}$};
 
    \draw[dashed] (2,1) circle (0.6) ;
    \draw[fill=black] (2,1) circle (0.05) ;
    
    \end{tikzpicture}
    \caption{Representation of a scatterer at a position $\textbf{r}_n$.}
    \label{fig:appendix_diffusion}
\end{figure}

Let us examine a general case where the scattering center is not necessarily at the origin, but at a position say $\textbf{r}_n$ as represented on the Fig.~\ref{fig:appendix_diffusion}. This case is not explicitly present in the literature the authors have consulted, except for Ref.~\cite{Starrett2016}, where the author might have simply used Bloch theorem. However, a proper derivation, which almost entirely follows some derivations where the scatterer is at the origin, see for example \cite{Mahan2000,Marchildon2002}, is given further.
Let us set the propagation vector $\textbf{k}$ in the $\textbf{z}$ direction and return to the Lippmann-Schwinger equation that was given in Eq.~(\ref{eq:lippmann2}), that we write for a single scatterer as,
\begin{equation*}
\psi(\textbf{r}) = e^{i\textbf{k}\cdot\textbf{r}} - \frac{1}{2\pi} \int_{\mathbb{R}^3}\frac{e^{ik|\textbf{r}-\textbf{r}_0|}}{|\textbf{r}-\textbf{r}_0|} V_0(| \textbf{r}_0-\textbf{r}_n |)\psi(\textbf{r}_0)\mathrm{d}^3r_0 .
\end{equation*}
If we set $\textbf{r} = \textbf{r}_n+\textbf{r}'$, to move in the scatterer's reference frame, and perform a variable change in the integral $\textbf{r}_1 = \textbf{r}_0 - \textbf{r}_n$, we get, 
\begin{equation*}
\psi(\textbf{r}'+\textbf{r}_n) = e^{i\textbf{k}\cdot\textbf{r}_n}e^{i\textbf{k}\cdot\textbf{r}'} - \frac{1}{2\pi} \int_{\mathbb{R}^3}\frac{e^{ik|\textbf{r}'-\textbf{r}_1|}}{|\textbf{r}'-\textbf{r}_1|} V_0(| \textbf{r}_1|)\psi(\textbf{r}_1+\textbf{r}_n)\mathrm{d}^3r_1 .
\end{equation*}
If we expand the Green function inside the integral, we find the asymptotic behavior of $\psi_n(\textbf{r}') \equiv \psi(\textbf{r}'+\textbf{r}_n) $ as being, 
\begin{equation*}
\psi_n(\textbf{r}') \underset{|\textbf{r}'| \to +\infty}{=}e^{i\textbf{k}\cdot\textbf{r}_n}e^{i\textbf{k}\cdot\textbf{r}'} - \frac{1}{2\pi} \frac{e^{ikr'}}{r'}\int_{\mathbb{R}^3} e^{-i\textbf{k}'\cdot \textbf{r}_1} V_0(| \textbf{r}_1|)\psi_n(\textbf{r}_1)\mathrm{d}^3r_1,
\end{equation*}
where $\textbf{k}' = k\textbf{r}'/r'$, is the direction of the scattered spherical wave. We see directly from the integral form, that asymptotically, the wave function writes $\psi_n(\textbf{r}') \underset{|\textbf{r}'| \to +\infty}{=} e^{i\textbf{k}\cdot\textbf{r}_n}e^{i\textbf{k}\cdot\textbf{r}'} + f(\theta')e^{ikr'}/r'$. Here $\theta'$ is the angle between $\textbf{r}'$ and the incoming $\textbf{k}$ vector, and $f$ is what is commonly called - the diffusion amplitude.

This wave form can be expanded on the basis of Legendre polynomials, using the expansion of $e^{i\textbf{k}\cdot\textbf{r}'}=e^{ikr'\cos(\theta')}$ We can write,
\begin{equation*}
\psi_n(\textbf{r}') \underset{|\textbf{r}'| \to +\infty}{=}e^{i\textbf{k}\cdot\textbf{r}_n}\sum_{\ell=0}^{+\infty} (2\ell+1)i^{\ell}j_\ell(kr')P_{\ell}(\cos(\theta')) + f(\theta')\frac{e^{ikr'}}{r'}.
\end{equation*}
In its asymptotic form, the Bessel function can be expressed as $j_\ell(kr') \underset{|r'| \to +\infty}{=} \sin(kr'-\pi\ell/2)/kr'$, thus giving us on the one hand,
\begin{equation*}
\begin{aligned}
\psi_n(\textbf{r}') \underset{|\textbf{r}'| \to +\infty}{=} &\frac{e^{ikr'}}{r'}\left[f(\theta')+\sum_{\ell=0}^{+\infty} e^{i\textbf{k}\cdot\textbf{r}_n}\frac{(2\ell+1)}{2ik}P_{\ell}(\cos(\theta')) \right]+\frac{e^{-ikr'}}{r'}\sum_{\ell=0}^{+\infty} e^{i\textbf{k}\cdot\textbf{r}_n}\frac{(2\ell+1)}{2ik}P_{\ell}(\cos(\theta')).
\end{aligned}
\end{equation*}
On the other hand, as explained in for example \cite{Marchildon2002}, with the spherically symmetric potential, the total wave function can be expanded in a general $\varphi'$ independent form as, 
\begin{equation*}
\psi_n(\textbf{r}') = \sum_{\ell=0}^{+\infty} c_\ell(2\ell+1)i^\ell\frac{u_\ell(r')}{r'}P_\ell(\cos(\theta')),
\end{equation*}
where $u_l$ is a solution of the radial Schr\"odinger equation, which writes $u_\ell''(r')+[k^2-2V(r')-\ell(\ell+1)/r'^2]u_\ell(r') = 0$. The asymptotic form of the $u_\ell$ function is $u_\ell(r')\underset{|r'| \to +\infty}{=} \sin(kr'-\pi\ell/2+\delta_\ell)/k$, where the $\delta_\ell(k)$ are the scattering phase shifts. Decomposing the sine with complex exponentials, we obtain as previously, 
\begin{equation*}
\begin{aligned}
\psi_n(\textbf{r}') \underset{|\textbf{r}'| \to +\infty}{=} &\frac{e^{ikr'}}{r'}\sum_{\ell=0}^{+\infty} \frac{c_\ell(2\ell+1)e^{i\delta_\ell}}{2ik}P_{\ell}(\cos(\theta'))+\frac{e^{-ikr'}}{r'}\sum_{\ell=0}^{+\infty} \frac{c_\ell(2\ell+1)e^{-i\delta_\ell}}{2ik}P_{\ell}(\cos(\theta')).
\end{aligned}
\end{equation*}
We can now identify the two expressions. We obtain first from the $e^{-ikr'}/r'$ part : $c_\ell e^{-i\delta_\ell} = e^{i\textbf{k}\cdot\textbf{r}_n}$. Identifying the other sum gives us finally the diffusion amplitude in the case of a single scatterer displaced from the origin to a position $\textbf{r}_n$ as being,
\begin{equation*}
f(\theta') = \frac{e^{i\textbf{k}\cdot\textbf{r}_n}}{k}\sum_{\ell=0}^{+\infty} (2\ell+1)e^{i\delta_\ell}\sin(\delta_\ell)P_{\ell}(\cos(\theta')).
\end{equation*}
This gives us also a derivation of the local approximation expressed in Evans' paper. Now that we have the expression of the coefficients $c_\ell$, we can approach the total wave function (plane wave and scattered spherical wave) as,
\begin{equation*}
\psi(\textbf{r}) \underset{|\textbf{r}-\textbf{r}_n| < r_{\mathrm{ws}}}{=} e^{i\textbf{k}\cdot\textbf{r}_n}\sum_{\ell=0}^{+\infty} (2\ell+1)i^\ell e^{i\delta_\ell}\frac{u_\ell(|\textbf{r}-\textbf{r}_n|)}{|\textbf{r}-\textbf{r}_n|}P_\ell(\cos(\theta')),
\end{equation*}
where $\theta'$ is the local angle between $\textbf{r}-\textbf{r}_n$ and $\textbf{k}$.
\section{Appendix C: Expression of the transport cross section}\label{appC}

Let us calculate the effective transport cross-section. We know that the scattering amplitude $f(\theta)$ reads
\begin{equation*}
f(\theta) = \frac{1}{p}\sum_{n=0}^{+\infty} (2n+1) e^{i\delta_n}\sin(\delta_n)P_n(\cos(\theta)),
\end{equation*}
where $P_n$ are the Legendre polynomials which satisfy the following identities \cite{Abramowitz1964}
\begin{equation*}
\int_{-1}^{1}P_n(x)P_m(x) dx = \frac{2}{2n+1}\delta_{mn},
\end{equation*}
as well as the following recursion relation
\begin{equation*}
(n+1) P_{n+1}(x) = (2n+1)xP_n(x)-nP_{n-1}(x).
\end{equation*}
We can now calculate the effective transport cross-section, which can be written as 
\begin{equation}\label{tra}
\begin{gathered}
\sigma_{tr}(p) = \int (1-\cos(\theta)) |f(\theta)|^2 d\Omega = \\ 2\pi\int_{0}^{\pi}(1-\cos(\theta))\sin(\theta) |f(\theta)|^2 d\theta = 2\pi \int_{-1}^{1} (1-x)|f(x)|^2dx.
\end{gathered}
\end{equation}
and expand the sum by noticing that $|f(x)|^2 = f(x)f^{*}(x)$, were $f^*$ denotes the complex conjugate of $f$. We get
\begin{align}
\sigma_{tr}(p) =& \frac{2\pi}{p^2} \sum_{n,m} (2n+1)(2m+1) e^{i(\delta_n-\delta_m)}\sin(\delta_n)\sin(\delta_m)\\
&\times\int_{-1}^{1}(1-x)P_n(x)P_m(x)dx.
\end{align}
From the relations satisfied by the Legendre polynomials, one obtains
\begin{equation*}
\begin{gathered}
\int_{-1}^{1}(1-x)P_n(x)P_m(x)dx = \frac{2}{2n+1}\delta_{mn}- \int_{-1}^{1} \left[\frac{n+1}{2n+1} P_{n+1}(x) + \frac{n}{2n+1}P_{n-1}(x)\right]P_m(x)dx,
\end{gathered}
\end{equation*}
and thus
\begin{equation}\label{leg}
\begin{gathered}
\int_{-1}^{1}(1-x)P_n(x)P_m(x)dx = \frac{2}{2n+1}\delta_{mn} - \frac{2n+2}{(2n+1)(2n+3)} \delta_{n+1,m} - \frac{2n}{(2n+1)(2n-1)} \delta_{n-1,m}.
\end{gathered}
\end{equation}
Inserting Eq.~(\ref{leg}) into expression (\ref{tra}) of the transport cross-section and changing the index as $n \rightarrow n+1$ for the last term (keeping the sum unchanged) yields
\begin{equation*}
\begin{gathered}
\sigma_{tr}(p) = \frac{4\pi}{p^2} \sum_{n=0}^{+\infty} (2n+1)\sin^2(\delta_n) - (n+1) e^{i(\delta_n-\delta_{n+1})}\sin(\delta_n)\sin(\delta_{n+1})\\ - (n+1) e^{i(\delta_{n+1}-\delta_{n})}\sin(\delta_{n+1})\sin(\delta_{n}).
\end{gathered}
\end{equation*}
Gathering the terms and resorting to the identity $e^{i(\delta_n-\delta_{n+1})} + e^{i(\delta_{n+1}-\delta_{n})} = 2\cos(\delta_{n+1}-\delta_{n})$, we can write
\begin{equation*}
\sigma_{tr}(p) = \frac{4\pi}{p^2} \sum_{n=0}^{+\infty} (2n+1)\sin^2(\delta_n) - (2n+2) \cos(\delta_{n+1}-\delta_{n})\sin(\delta_n)\sin(\delta_{n+1}).
\end{equation*}
The identity $\cos(\delta_{n+1}-\delta_{n}) = \cos(\delta_{n+1}) \cos(\delta_{n}) + \sin(\delta_{n+1}) \sin(\delta_{n})$ enables us to transform the sum into
\begin{equation*}
\begin{gathered}
\sigma_{tr}(p) = \frac{4\pi}{p^2} \sum_{n=0}^{+\infty} (2n+1)\sin^2(\delta_n) - (2n+2) \cos(\delta_{n})\cos(\delta_{n+1}) \sin(\delta_n)\sin(\delta_{n+1}) \\- (2n+2)\sin^2(\delta_n)\sin^2(\delta_{n+1}).
\end{gathered}
\end{equation*}
We have
\begin{equation*}
\begin{gathered}
\sin^2(\delta_{n+1}-\delta_{n}) = \sin^2(\delta_{n+1})\cos^2(\delta_{n}) - 2\sin(\delta_{n})\cos(\delta_{n})\sin(\delta_{n+1})\cos(\delta_{n+1}) + \cos^2(\delta_{n+1})\sin^2(\delta_{n}), 
\end{gathered}
\end{equation*}
and using $\cos^2 x=1-\sin^2x$, one obtains 
\begin{equation*}
\begin{gathered}
\sin^2(\delta_{n+1}-\delta_{n}) = \sin^2(\delta_{n+1}) + \sin^2(\delta_{n}) - 2\sin(\delta_{n})\cos(\delta_{n})\sin(\delta_{n+1})\cos(\delta_{n+1}) -2 \sin^2(\delta_{n+1})\sin^2(\delta_{n}), 
\end{gathered}
\end{equation*}
which is equal to 
\begin{equation*}
\begin{gathered}
\sigma_{tr}(p) = \frac{4\pi}{p^2} \sum_{n=0}^{+\infty} (n\sin^2(\delta_n) - (n+1)\sin^2(\delta_{n+1}) + (n+1)\sin^2(\delta_{n+1}-\delta_{n})).
\end{gathered}
\end{equation*}
The first two terms constitute a telescoping sum in which subsequent terms cancel each other, leaving only initial term, which is zero. Then, as required and replacing the summation index $n$ by $\ell$ to remain consistent with our notations:
\begin{equation*}
\sigma_{tr}(p) = \frac{4\pi}{p^2} \sum_{\ell=0}^{\infty} (\ell+1)\sin^2[\delta_{\ell+1}(p)-\delta_\ell(p)].
\end{equation*}

\section{Appendix D: Imaginary part of the conductivity using Kramers-Kronig relations}\label{appD}

Since we have to compute the integral of the right-hand side of Eq. (\ref{imag}), with the Drude conductivity
\begin{equation}
    \Re\left[\sigma(\omega)\right]=\frac{\sigma_0}{1+\tau_p^2\omega^2},
\end{equation}
where $\sigma_0$ is given in Eq. (\ref{sigma0}), we have 
\begin{equation*}
\Im\left[\sigma(\omega)\right]=-\frac{2\omega\sigma_0}{\pi}\mathcal {P}\!\!\int _{0}^{\infty}\frac{1}{(\omega'^{2}-\omega ^{2})(1+\tau_p^2\omega^2)}\,d\omega '.
\end{equation*} 
Let us therefore consider the integral
\begin{equation}
    I=\mathcal {P}\int_0^{\infty}\frac{d\omega'}{(\omega'^2-\omega^2)(1+\tau_p^2\omega'^2)}.
\end{equation}

We have
\begin{equation*}
\frac{1}{(\omega'^2-\omega^2)(1+\tau_p^2\omega'^2)}=\frac{1}{1+\tau_p^2\omega^2}\left[-\frac{\tau_p^2}{1+\tau_p^2\omega'^2}+\frac{1}{\omega'^2-\omega^2}\right]
\end{equation*}
and 
\begin{equation}\label{des}
I=\frac{1}{1+\tau_p^2\omega^2}\left[-\tau_p^2\int_0^{\infty}\frac{d\omega'}{1+\tau_p^2\omega'^2}+\mathcal{P}\int_0^{\infty}\frac{d\omega'}{\omega'^2-\omega^2}\right].
\end{equation}
We have
\begin{equation*}
\int_0^{\infty}\frac{d\omega'}{1+\tau_p^2\omega'^2}=\left.\frac{1}{\tau_p}\mathrm{arctan}(\tau_px)\right|_0^{\infty}=\frac{\pi}{2\tau_p}.
\end{equation*}
The second integral in Eq.(\ref{des}) is
\begin{equation*}
    \mathcal{P}\int_0^{\infty}\frac{d\omega'}{\omega'^2-\omega^2}=\frac{1}{\omega}\mathcal{P}\int_0^{\infty}\frac{d\omega'}{\omega'^2-1}
\end{equation*}
and
\begin{equation*}
\mathcal{P}\int_0^{\infty}\frac{d\omega'}{\omega'^2-1}=\lim_{\epsilon\rightarrow 0}\left[\int_0^{1-\epsilon}\frac{d\omega'}{\omega'^2-1}+\int_{1+\epsilon}^{\infty}\frac{d\omega'}{\omega'^2-1}\right].
\end{equation*}
Making the change of variable $\omega'\rightarrow 1/\omega'$ in the last integral of the right-hand side of the latter equation, we are left with
\begin{equation*}
\mathcal{P}\int_0^{\infty}\frac{d\omega'}{\omega'^2-1}=\lim_{\epsilon\rightarrow 0}\left[\int_0^{1-\epsilon}\frac{d\omega'}{\omega'^2-1}-\int_{0}^{\frac{1}{1+\epsilon}}\frac{d\omega'}{\omega'^2-1}\right].
\end{equation*}
or equivalently
\begin{equation*}
\mathcal{P}\int_0^{\infty}\frac{d\omega'}{\omega'^2-1}=\lim_{\epsilon\rightarrow 0}\int_{\frac{1}{1+\epsilon}}^{1-\epsilon}\frac{d\omega'}{\omega'^2-1},
\end{equation*}
which tends to 0 as $\epsilon\rightarrow 0$. Finally, we have
\begin{equation*}
    I=\mathcal {P}\int_0^{\infty}\frac{d\omega'}{(\omega'^2-\omega^2)(1+\tau_p^2\omega'^2)}=-\frac{\pi}{2}\frac{\tau_p}{1+\tau_p^2\omega^2}.
\end{equation*}
The imaginary part of the conductivity is thus
\begin{equation*}
\Im\left[\sigma(\omega)\right]=\frac{\omega\tau_p\sigma_0}{1+\tau_p^2\omega^2}.
\end{equation*}

\end{document}